\documentclass[aps,pra,onecolumn,showkeys,superscriptaddress,10pt,nofootinbib]{revtex4-2}
\usepackage{multirow,amsthm,amssymb,amsbsy,amsmath,epsfig,epstopdf,bm,float}

\usepackage{amsmath}
\usepackage{amssymb}
\usepackage{graphicx}
\usepackage{hyperref}
\usepackage{comment}
\hypersetup{
	colorlinks   = true, %Colours links instead of ugly boxes
	urlcolor     = darkgreen, %Colour for external hyperlinks
	linkcolor    = darkblue, %Colour of internal links
	citecolor   = darkred %Colour of citations
}
\usepackage{xcolor}
\colorlet{darkred}{red!85!black}
\colorlet{darkgreen}{green!50!black}
\colorlet{darkblue}{blue!60!black}

\def\Var{{\textrm{Var}}\,}
\usepackage{bm}

  \newcommand{\av}[1]{\left\langle#1\right\rangle}
  \newcommand{\cbr}[1]{\left(#1\right)}

  \newcommand{\mar}[1]{\textcolor{black}{#1}}
  \newcommand{\new}[1]{\textcolor{black}{#1}}

\providecommand{\keywords}[1]
{
  \small	
  \textbf{\textit{Keywords---}} #1
}

\begin{document}

\title{Thermalization in  high-dimensional systems: the (weak) role of chaos}
%\title{The weak role of chaos for the  thermalization  and statistical features in  high dimensional systems}

\author{Marco Baldovin}
\affiliation{Institute for Complex Systems, CNR, 00185, Rome, Italy}

\author{Marco Cattaneo}
\affiliation{
Department of Physics, University of Helsinki, P.O. Box 43, FI-00014 Helsinki, Finland}

\author{Dario Lucente}
\affiliation{Dipartimento di Fisica, Università di Roma “La Sapienza”, P.le Aldo Moro 5, 00185, Rome, Italy}

\author{Paolo Muratore-Ginanneschi}
\affiliation{ 
Department of Mathematics and Statistics, University of Helsinki, P.O. Box 68, FI-00014 Helsinki, Finland}

\author{Angelo Vulpiani}
\affiliation{Dipartimento di Fisica, Università di Roma “La Sapienza”, P.le Aldo Moro 5, 00185, Rome, Italy}
\affiliation{Institute for Complex Systems, CNR, 00185, Rome, Italy}


\begin{abstract}
%Statistical mechanics  has proved extremely powerful and effective in describing the macroscopic features of many-particle systems in very diverse contexts,  both for the equilibrium and the non equilibrium cases. However, there is still not be a general consensus about the reasons underlying this success. In particular, 
In their seminal work, Fermi, Pasta, Ulam and Tsingou explored the connection between statistical mechanics and dynamical properties, such as chaos and ergodicity. Even today, seventy years later, the topic is not fully understood: 
while most results of statistical mechanics require the ergodic hypothesis to be rigorously proved,  there are  many indications  that these predictions,  both in and out of equilibrium, hold even in the absence of a rigorous form of ergodicity.

Motivated by the above considerations, in this work we reconsider the point of view that the relevant ingredients for the validity of statistical mechanics are the large number of degrees of freedom and the choice of extensive observables, while the details of the dynamics do not play an essential role.
This is the idea  behind Khinchin's famous proof of the typicality of macroscopic observables at equilibrium. We extend this perspective to the context of non equilibrium, by investigating the thermalization properties of both harmonic (integrable) and nonharmonic (chaotic) oscillator chains initially prepared in out-of-equilibrium conditions.
%, i.e. within a set of initial conditions whose measure vanishes in the thermodynamic limit.
In integrable systems, thermalization occurs, or not, depending on the observable. In the chaotic regime, instead, thermalization is reached by any observable, although the relaxation timescale might be larger than the observation time.  
\end{abstract}

\maketitle
\section{Introduction}

The success of statistical mechanics (SM) in the
description of macroscopic systems is the object of an old debate:
the ability  of SM to describe the emergence of collective behaviours in systems
with a very large number of degrees of freedom is an unquestionable fact, but there is no general 
consensus  about the actual reasons behind it. Most results of SM are derived under strong mathematical assumptions on the nature of the dynamics (ergodic hypothesis), but these predictions have been shown to hold true also in cases where the requirements are not verified. On the other hand, the numerical experiment by Fermi, Pasta, Ulam and Tsingou (FPUT)~\cite{FPU55} has shown that even when the conditions for ergodicity are fulfilled, the time scales to reach the equilibrium predictions, if the system is not prepared in a thermal state, may be extremely large (suggesting that the relation between SM and dynamics may be more complex than originally supposed by Fermi himself). 

Dealing with   this  difficult  problem requires the careful consideration of many different aspects, both mathematical and physical:
from the role of the microscopic dynamics, to  the probability
description and the relevance of coarse-graining.
Answers that have been proposed in the literature roughly fall into three different classes. First, there is the point of view that  privileges the dynamical aspects, regarding the presence of chaos as the  basic   ingredient  for the validity
of SM~\cite{prigogine1979nouvelle,gaspard_1998,baldovin2025foundations}.
A second perspective proposes to found SM on the maximum entropy principle: this approach does not consider SM as a physical theory, but instead as an inference methodology based on incomplete information~\cite{J67}.
Finally, there is the approach that dates back to  the work of Boltzmann, and then mathematically
developed by Khinchin, where the main ingredients are recognized to be the
presence of a very large number of degrees of freedom, and the focus on extensive observables~\cite{lebowitz1993boltzmann,Khinchin1949}.

The aim of this work is not to enter into a general analysis of the foundation of SM: for more detailed discussions, the interested reader can refer to~\cite{zanghi2005fondamenti,da18}, as well as the recent review~\cite{baldovin2025foundations}. We will focus instead on the extension of the above problem to the case of relaxation from non-equilibrium conditions. To this aim, we will use as case studies both harmonic (integrable) and FPUT-like (chaotic) chains. We anticipate that, for certain observables, even in the harmonic case one can have full thermalization. Specifically, starting from an initial condition far from the thermal equilibrium, after a transient the observed quantity shows features in agreement with the  prediction of  equilibrium SM. Of course, this behaviour is not shown by \textit{all} observables. All the physical quantities that can be written as functions of the energies of normal modes, for instance, are exactly conserved by the dynamics; however this class of observables is somehow pathological and not very representative of the thermodynamic state of the chain. It is more interesting to observe that even extensive observables defined as the sum of single-particle functions, not immediately relatable to conserved quantities, do not \mar{always} thermalize. This apparent contradiction of Khinchin's point of view finds its explanation in the choice of the initial conditions: \new{the out-of-equilibrium initial states that we consider are interesting from the experimental point of view, but are neglected by Khinchin's argument, which focuses on equilibrium initial conditions}. 

By analyzing the effect of nonlinear terms in the dynamics, we further assess the role of chaos and ergodicity in this context. We show that these strong dynamical properties eventually ensure relaxation for all observables, even if the system is prepared in a ``pathological'' initial condition which would lead to ergodicity breaking in the harmonic limit. However, as already shown in the FPUT seminal work and subsequent literature, the relaxation times required to reach thermalization can be  extremely long. For all practical purposes, the inclusion of small nonharmonic corrections may be therefore irrelevant. 

%Nonetheless, their relevance for physical irreversibility appears to be limited.

%By analyzing the effect of nonharmonicity in the dynamics, we further assess the role of chaos and ergodicity in this context. We show that these strong dynamical properties effectively eliminate the observable dependence of thermalization, ensuring relaxation for all observables. Nonetheless, their relevance for physical irreversibility appears to be limited.
%We can  summarize the main result of the present paper as an attempt   try to extent the approach due to Khinchin~\cite{Khinchin1949} to  initial  conditions far from equilibrium.
%Let us emphasize that, within this framework, we allow for the existence of ``pathological'' initial conditions for which certain observables fail to thermalize. However, the measure of such states approaches $0$ in the thermodynamic limit, $N \to \infty$. By analyzing the effect of nonlinear terms in the dynamics, we further assess the role of chaos and ergodicity in this context. We show that these strong dynamical properties effectively eliminate the observable dependence of thermalization, ensuring relaxation for all observables. Nonetheless, their relevance for physical irreversibility appears to be limited.

The paper is organized as follows: 
in Sec.~\ref{sec:irreversible} we present some general results about the irreversibility
in paradigmatic models of macroscopic objects, discussing the mechanisms underlying the irreversible behavior, with particular emphasis on the role of initial conditions and the large number of degrees of freedom.
Sec.~\ref{sec:FPUT} is devoted to the analytical and numerical study of thermalization and energy equipartition in high dimensional systems of linear and weakly nonlinear oscillators. \mar{We first study a set of initial states for which interesting observables start far from equilibrium and then thermalize at late times, even in the harmonic case. Next, we} consider a class of far-from-equilibrium initial conditions which differs significantly from the standard FPUT setup, although it is deeply connected to this setup when expressed in a different representation (i.e. the lattice-site basis rather than normal modes). In this situation, we show that, in the harmonic cases, at long times the system relaxes to average values that are equal for all the degrees of freedom (i.e. equipartition is achieved), but these values do not coincide with those predicted by the microcanonical ensemble. For weakly non integrable systems, the presence of small nonlinearities restores the agreement of long-time averages and the prediction of SM, although the characteristic time-scale can be  very long. 
In Sec.~\ref{sec:conclusion} we draw our conclusions. The Appendices contain the detailed analytical derivations supporting our results.

\section{Physical irreversibility  in  high-dimensional systems: the role of chaos}\label{sec:irreversible}

Let us consider a system with $N \gg 1$ degrees of freedom, 
  initially prepared ``far enough from equilibrium'',  i.e. in a condition such that
  \begin{equation*}
M(0)= \langle M \rangle_{eq} +\delta M(0) \quad  \quad  \text{with} \quad \quad   |\delta M(0)| \gg \sigma_M \, ,
  \end{equation*}
 for  some  macroscopic variable $M$ 
 depending on all  microscopic variables (or, at least, a finite fraction of them), being $\langle M \rangle_{eq}$ the equilibrium average of the observable $M$, and $\sigma_M$ the amplitude of its typical fluctuations at equilibrium. By ``physical  irreversibility'' or ``thermalization'' we mean that, in a single realization, $M(t)$ approaches the equilibrium value $\av{M}_{eq}$ and stays close to it for all times $t > \tau_R$ (where $\tau_R$ is some typical relaxation time).
Let us stress already at this stage  an important  conceptual point:  physical  irreversibility
involves a  single  system with a  large number of particles~\cite{baldovin2025foundations,sarracino2025nonequilibrium}, not an average over a set of initial conditions. Unlike some  naive  presentations seem to suggest, 
statistical ensembles do not play any role in physical  irreversibility. 
 It is therefore  important to underline   the   deep    difference  between the physical   irreversibility
in a single (macroscopic) object, and the relaxation of a phase-space probability 
distribution $\rho({\bf X},t)$ to an invariant  distribution, which is a relevant but different problem~\cite{cerino2016role,baldovin2025foundations,lucente2025conceptual}.
The relaxation to the invariant measure is a
mathematical  property of an  ensemble of the initial conditions:
 independently of the initial density distribution  $\rho({\bf X},0)$,
 if the system is mixing, 
 for large $t$   one  has  $ \rho({\bf X}, t) \,\, \to \rho_{inv}({\bf X})$.
This  is  for sure an important aspect, especially in the dynamical
systems context, but it is not related to  irreversibility in the sense commonly attributed to it. When a cup of hot coffee cools down to room temperature, the phenomenon does not involve an ensemble of cups initially prepared in out-of-equilibrium conditions: it is already manifest at the level of the single cup.

In the rest of this section we will review some examples of macroscopic irreversibility, discussing the properties of the underlying dynamics.

  \subsection{The $H$-theorem}
 While a general mathematical understanding of the necessary conditions for physical irreversibility is still missing, for some systems rigorous results can be established by considering suitable scaling limits.
 The  archetypal  example of  approach to equilibrium  is the celebrated $H$-theorem
 for diluted gases. This  result has been proved in a rigorous way in the so called Grad-Boltzmann limit, for sufficiently small times~\cite{LANFORD198170}, and recently extended to arbitrary large times under some regularity assumptions on the solution of the Boltzmann equation~\cite{deng2024long}. Similar results can be derived in the context of simplified models where the technical difficulties are much less severe than in the case considered by Boltzmann. Two notable examples are the Kac ring~\cite{kac1956some,gottwald09} and the Ehrenfest model~\cite{eh56}: they both describe the evolution of a set of binary variables $\{\sigma_i\}_{i=1}^N$, the former with a deterministic map, the latter with a stochastic dynamics. By defining the variable $M(t)$ as the number of positive $\sigma_i$ at time $t$,  it can be shown that, if $N$  is large,  the single realization of $M(t)$ not only thermalizes, but is typically very close  to some $\langle M(t) \rangle$~\cite{gottwald09,baldovin2019irreversibility}. The average has different meanings in the two systems: it is an average over the disorder in the Kac model, an average over the realizations in the Ehrenfest one.  
 From a conceptual point of view, the most important properties of the examples mentioned above are the large number of degrees of freedom and the atypical initial conditions.
 Indeed, the dynamical properties of these models are profoundly different: the Kac ring dynamics is periodic (integrable), the Ehrenfest model is an ergodic Markov chain, and collisions among particles—together with the resulting chaotic behavior—constitute the basic physical mechanism underlying the H-theorem. In the following, we describe a different dynamical mechanism, namely the dephasing, for the emergence of irreversibility in high-dimensional integrable systems. 
 %From a mathematical perspective the three microscopic dynamics are instead quite different. In the Ehrenfest model, both the microscopic variables $\{\sigma\}_{i=1}^N$ and the macroscopic observable $M$ are stochastic processes evolving as Markov chains. The microscopic dynamics of the Kac ring and of the hard-sphere for dilute gases are instead deterministic and time reversible, and the irrerversible behavior in both cases emerges in an appropriate scaling limit~\cite{gottwald09,pulvirenti2021brief}.
 %Still the physical mechanism leading to irreversibility is profoundly different in the two models, since the Kac ring dynamics is periodic while it is well known that the collision among particles and the resulting chaotic behavior is the basic physical mechanism  in the H theorem.
 %From these examples it  is not completely clear what is the precise role of chaos in determining irreversible behavior for macroscopic observables. In the following we will see how in integrable systems one can have an irreversible behaviour given by a very different mechanism.

% As noted by Khinchin, the basic ingredients for explaining physical irreversibility relies on the large number of degrees of freedom and  
  %It is well known that the collision among particles  is the basic physical mechanism  in the H theorem,
  %in the following we will see how in integrable systems one can have an irreversible behaviour
  %given by a very different mechanism.
  
\subsection{ Chaotic and non chaotic  models for a piston}

Consider  a channel containing $N$ particles of mass $m$, closed by a fixed vertical wall on the left, and by a frictionless mobile wall of mass  $M$ on the right (the piston). 
Denoting by $x_n(t)$ the horizontal  coordinate of the $n-$th particle and by $X(t)$ the position of the piston,  one has therefore
$0\le x_n \le X$.
A constant force $-F$  acts on the piston and compresses the gas. Interactions are present  
among  the  particles inside the channel, as well as  between the particles and the walls. The total Hamiltonian  reads:
$$
H= {P^2 \over 2M} +\sum_ i {p^2_i \over 2m} +\sum_{i<j}U(|{\bf q}_i -{\bf q}_j|) +
U_w({\bf q}_1,..., {\bf q}_N, X) +F X \, ,
$$
where $U$ is the interaction potential among the particles, and $U_w$ denotes the interaction of the particles with the piston. By imposing reflecting boundary conditions at the origin, we model the presence of a fixed wall.
In the case of non interacting particles, in which $U=0$ and $U_w$ describes elastic collisions, the dynamics is not chaotic.
The equilibrium SM of such a system 
is trivial, therefore one can   find 
the equilibrium  position of the piston, $\langle X \rangle_{eq}$, and its variance $\sigma_X^2$. 
  %%%%%%%%%%%%%%%%%%%%%%%%  FIG.   %%%%%%%%%%%%%%%% BEGIN
\begin{figure}[t!]
\centering
\includegraphics[draft=false, scale=0.72, clip=true]{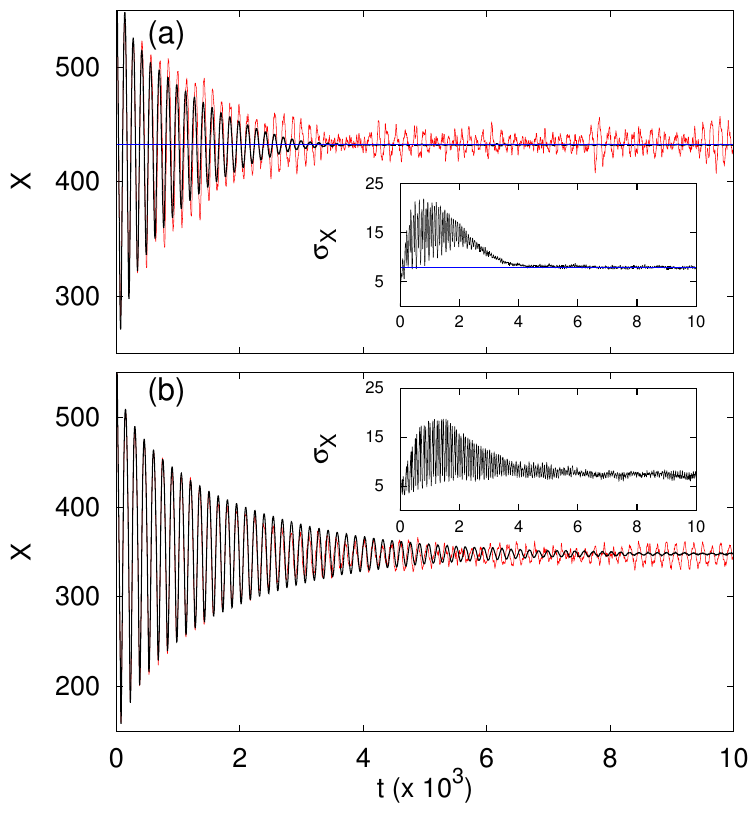}
\caption{$X(t)$ vs. $t$ for $N=1024$ and  $X(0)= X_{eq}+10 \sigma_{eq}$ in a chaotic piston (a), and in a non chaotic piston (b).
Red lines represent $X(t)$ for a single realization; black lines refer to the ensemble average $\langle X(t) \rangle$. Figure reproduced from~\cite{cerino2016role}.
}
\label{fig:cerino}
\end{figure}
%%%%%%%%%%%%%%%%%%%%%%%%  FIG. Typ-2  %%%%%%%%%%%%%%%% 

In the presence of interactions, e.g. with
 potentials like  $ U(r)=U_0/r^{12}$ and   $U_w=U_0\sum_n  |x_n-X|^{-12}$,
numerical computations show that chaos is present (namely  the first Lyapunov exponent is positive).
It is not possible to determine  analytically the equilibrium statistical properties, which can,
however, be studied numerically~\cite{cerino2016role}.

 Figure~\ref{fig:cerino} shows the behaviour of    the piston: when we prepare the system very far from thermal equilibrium, we observe damped oscillations around the equilibrium position.  Two remarks are in order. First, it should be noticed that
  the individual trajectories  are typical, i.e. close to the average. Second, the qualitative behaviour of the piston  is the same 
  for both the chaotic and the non chaotic cases:
 even far from equilibrium, relative fluctuations from the average trajectory  are small.

One may wonder about the dynamical mechanism causing the irreversible behaviour
of the piston.
Let us note that,  even  in the case of non interacting particles,
the collision of the piston with the gas induces an indirect interaction among the particles, which brings about a sort of effective randomization. The motion of the large mass $M$  is, therefore, the result of  many  weak dephased collisions.

\subsection{ Dephasing mechanism in integrable chains}
Consider a one-dimensional chain of $N$ particles of unit mass, characterized by first-neighbour interactions through a conservative force. The Hamiltonian of the model reads 
\begin{equation*}
\label{eq:Hamiltonian-generic-Chain}
H= \sum_{i=1}^{N} \left[  { p_i^2 \over 2  } + V\bigl( q_{i+1} - q_i \bigr)\right]\,,
\end{equation*}
where $V(r)$ is the potential. Many important physical models fall in this class: for instance, the original FPUT model is of this form, with $V(r)$ a $4-$th degree polynomial.

For other choices of the potential $V(r)$ the Hamiltonian system can be integrable: two notable examples are the harmonic case ($V(r)\propto r^2$) and the Toda chain ($V(r)\propto e^{-r}+r-1$). The action-angle coordinates $\{\mathcal{I}_i,\phi_i\}_{i=1}^N$ evolve in these cases as 
\begin{equation}
\nonumber
\begin{aligned}
    \mathcal{I}_i(t)&=\mathcal{I}_i(0)\\
    \phi_i(t)&=\phi_i(0)+\omega_i(\mathcal{I}_1,\cdots,\mathcal{I}_N) t
\end{aligned}
\end{equation}
%\{\mathcal{I}_i(t),\phi_i(t)\}_{i=1}^N=\{\mathcal{I}_i(0),\phi_i(0)+\omega_i(\mathcal{I}_1,\cdots,\mathcal{I}_N) t\}_{i=1}^N\,,    
for $1 \le i \le N$, with $\omega_i(\mathcal{I}_1,\cdots,\mathcal{I}_N)=\frac{\partial H}{\partial \mathcal{I}_i}$. Thus, the motions are confined to the tori specified by the initial values $\{\mathcal{I}_i(0)\}_{i=1}^N$of the action variables. 
%In spite of the apparent simplicity,  one can show that, in the limit of large $N$, some nontrivial  statistical features hold~\cite{Baldovin2021,baldovin2025foundations}.
In the rest of the section we focus only on the harmonic chain, but similar results have also been obtained in the context of the Toda model~\cite{Baldovin2021}.

\begin{figure}[t!]
\centering
\includegraphics[draft=false, scale=0.82, clip=true]{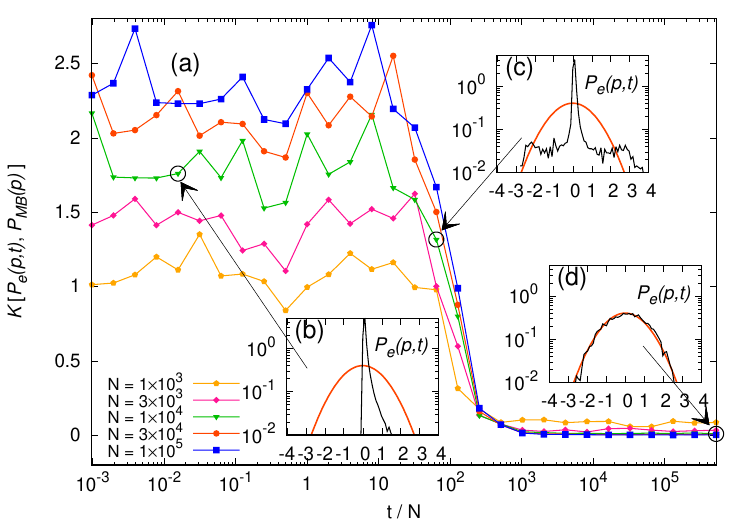}
\caption{ Kullback–Leibler divergence between the empirical distribution of the momenta  and the Maxwell–Boltzmann,
 as a function of time.
 Panel a) shows the behavior of  $K[P_e(p,t)| P_{MB}(p)]$
 for different values of $N$.
 The system is initialized in a far-from-equilibrium state where only a small fraction of the modes is excited. 
 The convergence  to equilibrium takes place on a time scale that is proportional to $N$.
 Panels  b), c) and d)  show the detail of the empirical distribution $P_e(p,t)$ 
 at different times, for the $N=10^4$   case. Figure reproduced from~\cite{Baldovin2023}.
 }
\label{Fig:3}
\end{figure}

Consider a system with $N \gg 1$, and
an initial condition very far from thermal equilibrium. Specifically, we pick momenta $\{ p_n(0) \}_{n=1}^N$ according to a probability density $P_e(p,0)$ different from the Maxwell-Boltzmann distribution: $P_{MB}(p) \propto \exp\cbr{ - p^2 / 2mk_BT}$.
Let us note that $P_e(p,0)$  is not related to statistical ensembles: it can be considered as the empirical density, i.e. the one obtained, via a histogram,  from the values
 $(p_1(0), p_2(0),..., p_N(0))$.
In the linear case, the angles $\{\omega_i\}_{i=1}^N$ are constant, and we can analytically determine the evolution of the momenta $\{ p_n(t) \}$ (see Sec.~\ref{sec:linear-case} for explicit expressions). Therefore, in the limit $N\gg 1$, it is also possible  to compute the moments of $P_e(p,t)$~\cite{Baldovin2023}. 
Following the seminal paper by Fermi et al.~\cite{FPU55}, we can study the case in which the first $N^{*}$ normal modes are excited with equal energy $E$, that is
\begin{equation}
\nonumber
\mathcal{I}_i(0)= \begin{cases}
\frac{E}{\omega_i} & \quad i=1,.., N^*\\
 0 & \quad i> N^*    
\end{cases}
\end{equation}
where $1 \ll N^* \ll N$. A way to characterize the distance between the empirical density $P_e(p,t)$
 and the Maxwell–Boltzmann distribution is via the Kullback-Leibler divergence, see e.g. \cite[\S~2.3]{CoTh2006}:
 $$
  K[P_e(p,t)| P_{MB}(p)]= - \int P_e(p,t) \ln \Big( {  P_{MB}(p) \over P_e(p,t)} \Big)   \, dp\,.
 $$
Panel (a) of Fig.~\ref{Fig:3} shows how the values of the Kullback–Leibler divergence numerically computed for different $N$ tend to zero when $t\to\infty$. Panels (b-d) of Fig.~\ref{Fig:3} illustrate how the empirical density $P_e(p,t)$ relaxes to equilibrium at selected representative times for $N=10^4$. 
The  characteristic time for the convergence to the thermal equilibrium
is observed to depend linearly on $N$.
 %%%%%%%%%%%%%%%%%%%%%%%%  FIG.   %%%%%%%%%%%%%%%% BEGIN
%%%%%%%%%%%%%%%%%%%%%%%%  FIG. Typ-2  %%%%%%%%%%%%%%%% 

A simple argument suggests that the typical time necessary for dephasing  is $O(N/\omega_0)$ for $ N\gg 1$ and $N^*/N \ll 1$, in agreement with numerical computations: the modes $i$ and $j$ dephase if $\cos(\omega_{i}t)$ and $\cos(\omega_{j}t)$ behave as independent variables.
This can be expected for $|\omega_{i}-\omega_{j}|t >  c\, \pi$ where $c$ is some constant.
To estimate the typical relaxation time $T$, we need to find the minimum value of $|\omega_{i}-\omega_{j}|$
among  the possible pairs  $(i, j)$.  Considering the explicit formula for $\omega_{i}$ (see Sec.~\ref{sec:linear-case}) we find $T \sim N/\omega_0$~\cite{Baldovin2023}.

A comment is in order. The irreversible  behaviour of integrable systems such as the harmonic and the Toda chain may look similar to that of chaotic systems,  e.g.  hard sphere or systems of interacting particles with a generic non harmonic potential.
The resemblance is, however, only superficial. As already mentioned, in the case of  hard spheres irreversibility originates by the collisions among
different particles,  which  exchange energy and momentum, and leads to a chaotic dynamics. In the integrable systems  considered above, the
 thermalization is due to a dephasing mechanism connected to the properties of sums of trigonometric functions~\cite{kac1943distribution}.

\section{The role of chaos for thermalization, equilibration, and energy equipartition in the FPUT model}\label{sec:FPUT}

We consider the Hamiltonian of the well-known FPUT model with $N$ particles with unit mass and fixed boundary conditions \cite{Berman2005,Gallavotti2008,Benettin2011}, and a minor modification:
\begin{equation}
    \label{eqn:HamFPUT}
    H_{\alpha,\beta}(p,q)= \frac{1}{2}\sum_{j=1}^N p_j^2+\sum_{j=0}^N V_{\alpha,\beta}(q_j-q_{j+1})+\sum_{j=1}^N \frac{k_R}{2} q_j^2,\quad q_0=q_{N+1}=0,
\end{equation}
%with potential energy
where
\begin{equation}
    \label{eqn:PotentialFPUT}
    V_{\alpha,\beta}(r)=\frac{r^2}{2}+\alpha \frac{r^3}{3}+\beta \frac{r^4}{4}.
\end{equation}
In contrast to the standard FPUT Hamiltonian, we have included a uniform harmonic restoring term proportional to $k_R> 0$ along the entire chain. The presence of this term does not affect the conceptual conclusions of our work, while it should be considered as a technical ingredient that allows us to define a set of well-motivated local observables at each site. We further elaborate on this point later.

In what follows, we separately analyze the linear and non-linear cases. The linear case corresponds to $\alpha=\beta=0$.

\subsection{Linear case}\label{sec:linear-case}
If $\alpha=\beta=0$, we can recast the Hamiltonian of the FPUT model into block-quadratic form:
\begin{equation}
\label{eqn:HamQuadraticForm}
H_{0,0}(p,q) =\frac{1}{2} \mathbf{p}^T \mathbf{p} + \frac{1}{2} \mathbf{q}^T \mathsf{M}\, \mathbf{q},
\end{equation}
where $\mathbf{p}=(p_1,\ldots,p_N)^T$, $\mathbf{q}=(q_1,\ldots,q_N)^T$ and $\mathsf{M}$ is a $N\times N$ matrix defined through 
\begin{equation}
    \mathsf{M}_{jk} =\begin{cases}
         2 + k_R & \text{ if }j=k,\\
         -1 & \text{ if }j=k\pm 1,\\
         0 & \text{ otherwise.}
    \end{cases}
\end{equation}

We can diagonalize $\mathsf{M}$ using the orthogonal matrix $\mathsf{C}$ with entries
\begin{equation}
    \label{eqn:eigenvectors}
    C_{jk} =\sqrt{ \frac{2}{N+1}} \sin\left(\frac{\pi j k}{N+1}\right),\quad j,k=1,\ldots,N,
\end{equation}
As a consequence, we can write the Hamiltonian of the chain as the sum of $N$ free particle Hamiltonians:
\begin{equation}
\label{eqn:HamLinDiagonalized}
H_{0,0}(p,q) =\frac{1}{2} \sum_{j=1}^N \left(P_j^2+\omega_j^2 Q_j^2\right),
\end{equation}
where we have introduced the \textit{normal modes} 
\begin{equation}\label{eqn:fromNormalModesToLocal}
    Q_j = \sum_{k=1}^N C_{kj} q_k ,\quad P_j= \sum_{k=1}^N C_{kj} p_k,
\end{equation}
while 
\begin{equation} \label{eqn:frequenciesNM}
    \omega_j = \sqrt{4 \sin^2\left(\frac{\pi j}{2(N+1)}\right)+k_R},\quad j=1,\ldots,N,
\end{equation}
are the frequencies of the normal modes.

The energy of the $j$th normal mode is 
\begin{equation}\label{eqn:energyNM}
    E_j^\text{nm} = \frac{1}{2}(P_j^2+\omega_j^2 Q_j^2). 
\end{equation}
The energy of each normal mode is a conserved quantity of the system, and the quadratures evolve as  
\begin{equation}
\label{eqn:solutionLinear}
    Q_j(t) = A_j \cos(\omega_j t +\phi_j),\qquad P_j(t) = -\omega_j A_j \sin(\omega_j t +\phi_j),
\end{equation}
The amplitude and the phase of each quadrature encode the initial conditions:
\begin{equation}
\label{eqn:initialCond}
    A_j= \sqrt{Q_j(0)^2+(P_j(0)/\omega_j)^2},\qquad \phi_j=\arctan\left(-\frac{P_j(0)}{\omega_j Q_j(0)}\right).
\end{equation}

In the original FPUT problem \cite{FPU55}, the authors looked for energy equipartition between the normal modes, after switching on the non-linear interaction ($\alpha\neq 0$ and/or $\beta\neq 0$). In this work, we focus instead on a different set of physical quantities, namely the local ``on-site'' energies of the physical particles: 
\begin{equation}\label{eqn:energyOS}
    E_j^\text{os} = \frac{1}{2}(p_j^2 + k_R q_j^2)= K_j^\text{os}+U_j^\text{os}, 
\end{equation}
where we have introduced the kinetic $K$ and potential $U$ contributions. Note that these observables would not be fully meaningful from a physical perspective without the harmonic restoring force in \eqref{eqn:HamFPUT}. Moreover, we assume $k_R =\mathcal{O}(1)$, avoiding scenarios in which $k_R$ is a perturbative parameter or scales with $N$. 

We focus on this set of quantities for the sake of comparing the linear (integrable) versus non-linear (chaotic) regime. In the integrable scenario, the energy of each normal mode does not evolve in time, while it can properly thermalize in the chaotic regime (see the discussions in \cite{Berman2005,Gallavotti2008,Benettin2011}). In contrast, the on-site particle energies exhibit a rich phenomenology in both the integrable and chaotic case. 

To be more precise, we study the time average of the observables we are interested in. If $f(t)$  is an observable, we focus on 
\begin{equation}
    \label{eqn:timeAv}
    \overline{f}_t = \frac{1}{t}\int_0^t ds\, f(s). 
\end{equation}
In particular, we are interested in the infinite-time limit of this average, $\overline{f}_\infty$ defined for $t\rightarrow\infty$. In this way, we can study the thermalization regime. 
In the original FPUT paper \cite{FPU55}, the authors studied the evolution of $\overline{E_j^\text{nm}}_t$, including the long-time limit. In this work, we focus instead on $\overline{E_j^\text{os}}_t$.

\subsubsection{Long-time limit}
Here, we summarize some of the results we presented in a recent paper \cite{Cattaneo2025thermalization} and put them in the perspective of the problems of thermalization, energy equipartition, and equilibration in the FPUT model. We will refer to \cite{Cattaneo2025thermalization} for most of the formal proofs, while we will try to explain the reasoning behind our results in a heuristic way. 

The long time limit of the on-site energies in \eqref{eqn:energyOS} is \cite{Cattaneo2025thermalization}
\begin{equation}
    \label{eqn:longTimeEnergyOS}
    \overline{E_j^\text{os}}_\infty = \sum_{k=1}^N \frac{C_{jk}^2A_k^2}{4}(\omega_k^2+k_R).
\end{equation}
This formula holds under the assumption that the frequencies of the normal modes are nondegenerate, which is indeed satisfied by \eqref{eqn:frequenciesNM}.

Remarkably, Eq.~\eqref{eqn:longTimeEnergyOS} has a weak dependence on $j$, which appears only in the coefficients $C_{jk}$. The initial conditions of the motion are encoded in the amplitudes $A_k$ in \eqref{eqn:initialCond}. Moreover, fixing $j$ and considering $C_{jk}$ as a vector, the latter is delocalized over many sites from $1$ to $N$. In other words, from \eqref{eqn:fromNormalModesToLocal} we observe that the excitation of a local mode spreads over many normal modes. This fact can be properly quantified by the inverse participation ratio (IPR) of each vector \cite{Kramer1993},
\begin{equation}
    \text{IPR}_j = \sum_{k=1}^N C_{jk}^4. 
\end{equation}
The IPR is a measure of the delocalization of a vector and of how its components scale with $N$. If the vector is perfectly localized, i.e., if there exists only one $k$ for which $C_{jk}\neq 0$, then $\text{IPR}_j=1$. On the contrary, if the vector is perfectly delocalized, which means $C_{jk}^2=1/N$ for all $k$, then $\text{IPR}_j=1/N$.  

It can be shown (see Appendix~\ref{sec:IPR_proof}) that the IPR of the eigenvectors of the quadratic form $\mathsf{M}$, i.e., $C_{jk}$ fixing $j$, is given by 
\begin{equation}\label{eqn:IPR_1dChain_FBC}
    \text{IPR}_j = \frac{3}{2(N+1)} \text{ for all }j.
\end{equation}
Therefore, for $N\gg 1$ the normal modes are delocalized on the lattice, and their components scale with $N$ as those of a vector maximally delocalized. 

\subsubsection{Typicality of thermalized long-time averages in the microcanonical ensemble}

The on-site energies of the physical particles are observables that depend on several normal modes. In the spirit of Khinchin \cite{Khinchin1949}, we want to study the behavior of these quantities in the microcanonical ensemble. Specifically, we will show that, in the limit $N\gg 1$, \textit{almost all} states in the microcanonical energy shell display the same values of long-time averages of the on-site energies. 

Since our aim is to study the limit $N\gg 1$, we will perform all calculations in the canonical ensemble at temperature $T$ and employ them for the microcanonical ensemble also, using the well-known equivalence of ensembles in the thermodynamic limit \cite{huang2008statistical}. For a collection of harmonic oscillators, we set 
\begin{equation}\label{eqn:temperatureEnergyconversion}
    T = \frac{E}{N},
\end{equation}
where $E$ is the total energy of the system and we fix $k_B=1$. Averages over the canonical ensemble on the phase space, characterized by the probability distribution
\begin{equation}\label{eqn:canonicalDistribution}
    \rho_\text{can}(T)= \frac{e^{- H_{0,0}(p,q)/T}}{Z}, \text{ with }Z= (2\pi T)^N \frac{1}{\omega_1\ldots\omega_N},
\end{equation}
tend to averages over the microcanonical energy shell $E$ for $N\gg 1$. 

For the quadratures of the normal modes, the canonical average immediately reads
\begin{equation}\label{eqn:canonAVPQnormal}
    \langle P_j\rangle_\text{can}=\langle Q_j\rangle_\text{can}=0,\quad \langle P_j^2\rangle_\text{can} = T,\quad \langle Q_j^2\rangle_\text{can} = \frac{T}{\omega_j^2},\quad\langle P_j^4\rangle_\text{can} = 3T^2,\quad \langle Q_j^4\rangle_\text{can} = \frac{3T^2}{\omega_j^4} . 
\end{equation}
Then, a quick calculation yields
\begin{equation}\label{eqn:canonAVpqOnsite}
    \langle p_j^2\rangle_\text{can} = \sum_{k=1}^N C_{jk}^2 \langle P_k^2\rangle_\text{can} = T,\quad  \langle q_j^2\rangle_\text{can} = \sum_{k=1}^N C_{jk}^2 \langle Q_k^2\rangle_\text{can} =T \sum_{k=1}^N \frac{C_{jk}^2}{\omega_k^2},
\end{equation}
and
\begin{equation}\label{eqn:canonicalOnSiteEnergies}
    \langle E_j^\text{os}\rangle_\text{can} = \frac{T}{2}\left(1 + k_R \sum_{k=1}^N  \frac{C_{jk}^2}{\omega_k^2}\right).
\end{equation}

Next, we focus on the long-time average of the on-site energies, Eq.~\eqref{eqn:longTimeEnergyOS}. We can calculate the canonical average of  \eqref{eqn:longTimeEnergyOS} over the phase space of initial conditions $Q_k(0)$ and $P_k(0)$. We obtain
\begin{equation}
    \label{eqn:averageTime_canonical}
    \langle \overline{E_j^\text{os}}_\infty\rangle_\text{can} = \langle E_j^\text{os}\rangle_\text{can}, 
\end{equation}
as expected. Next, we compute the variance of the time-average over the canonical ensemble: 
\begin{equation}\label{eqn:calculationEj}
    \langle \overline{E_j^\text{os}}_\infty^2\rangle_\text{can} = \frac{1}{16} \sum_{k,k'=1}^N C_{jk}^2 C_{jk'}^2 \langle (Q_k^2+P_k^2/\omega_k^2)(Q_{k'}^2+P_{k'}^2/\omega_{k'}^2)\rangle_\text{can} (\omega_k^2 +k_R) (\omega_{k'}^2+k_R).
\end{equation}
If $k\neq k'$, in the above expression the canonical average factorizes: 
\begin{equation}
 \langle (Q_k^2+P_k^2/\omega_k^2)(Q_{k'}^2+P_{k'}^2/\omega_{k'}^2)\rangle_\text{can}=\langle (Q_k^2+P_k^2/\omega_k^2)\rangle_\text{can}\langle(Q_{k'}^2+P_{k'}^2/\omega_{k'}^2)\rangle_\text{can}   \,.
\end{equation}
\mar{Using the delocalized property of $C_{jk}^2$ captured by the IPR in \eqref{eqn:IPR_1dChain_FBC}, we can then argue that all terms in the double summation in~\eqref{eqn:calculationEj} are of the same order. Then, we can replace the $N$ terms with $k=k'$, which are far fewer than the terms $k\neq k'$, with
\begin{equation}
    C_{jk}^4 \langle (Q_k^2+P_k^2/\omega_k^2)\rangle_\text{can}^2 (\omega_k^2+k_R)^2
\end{equation}
in~\eqref{eqn:calculationEj}.
We refer to page 10 of the supplemental material of \cite{Cattaneo2025thermalization} for a formal proof of this point.} Finally, we obtain:
\begin{equation} \label{eqn:KhinchinOnSiteEnergies}
    \langle \overline{E_j^\text{os}}_\infty^2\rangle_\text{can} = \langle \overline{E_j^\text{os}}_\infty\rangle_\text{can}^2\left(1+\mathcal{O}\left(\frac{1}{N}\right)\right), \quad  \frac{\Var[\overline{E_j^\text{os}}_\infty]_\text{can}}{\langle \overline{E_j^\text{os}}_\infty\rangle_\text{can}^2} = \mathcal{O}\left(\frac{1}{N}\right)\rightarrow 0 \text{ for }N\rightarrow\infty.
\end{equation}
Therefore, if we sample the initial conditions $\xi$ in the microcanonical energy shell $\Omega_{E}$ with a uniform probability measure $\operatorname{Pr}$, the equivalence of statistical ensembles and the Chebishev inequality allow us to conclude \cite{Khinchin1949,baldovin2025foundations}
\begin{align}
\operatorname{Pr}\left( \xi \in \Omega_{E} : \frac{| \overline{E_j^\text{os}}_\infty-\langle E_j^\text{os}\rangle_\text{can}|}{\langle E_j^\text{os}\rangle_\text{can}}>\varepsilon\right) \rightarrow 0 \text{ for }N\rightarrow\infty
\label{eqn:nontypical}
\end{align}
for any $\varepsilon>0$.
%Therefore, the probability of drawing an initial state  with uniform distribution over the microcanonical energy shell such that its long-time average of the on-site energies will converge towards their microcanonical average goes to 1 for $N\gg 1$. 
In other words, long-time thermalization of the on-site energies is typical over the microcanonical energy shell of long harmonic chains. 

We would like to emphasize that this result is not limited to static scenarios, i.e., situations where \mar{macroscopic quantities that depend on} the on-site energies at $t=0$ are already at thermal equilibrium. To verify this, we consider the on-site energy of the half chain (we assume for simplicity that $N$ is even): 
\begin{equation} \label{eqn:half_energy}
    E^\text{os}_\text{half} = \sum_{j=1}^{N/2} E_j^\text{os}.
\end{equation}
As this is a macroscopic quantity that scales as $N$, time fluctuations for any $t\gg 1$ vanish as $N\rightarrow\infty$ \cite{Cattaneo2025thermalization}. It is therefore easier to study the relaxation to the equilibrium value. Moreover, we also separately consider the kinetic and potential components of \eqref{eqn:half_energy}, which we denote by $K^\text{os}_\text{half}$ and $U^\text{os}_\text{half}$, respectively.

We now analyze a scenario in which all momenta of the normal modes are initially set at the same value, while the positions start at 0:
\begin{equation}\label{eqn:initialConditionLinearCase}
    P_j(0)= \sqrt{2 T},\quad Q_j(0)=0 \quad \text{for all }j.
\end{equation}
The reader can immediately verify that, for this initial condition, 
\begin{equation} \label{eqn:theoPredictionHalfChain}
    \overline{K^{\text{os}}_j}_\infty = \langle K^{\text{os}}_j\rangle_\text{can} = \frac{T}{2 },\quad  \overline{U^{\text{os}}_j}_\infty = \langle U^{\text{os}}_j\rangle_\text{can} = \frac{T k_R}{2} \sum_{k=1}^N\frac{C_{jk}^2}{\omega_k^2}.
\end{equation}
So, the on-site energy of the half chain together with its kinetic and potential components thermalize for $N\gg 1$, provided that fluctuations tend to zero as time increases. Remarkably, these observables at $t=0$ are far from equilibrium. For instance, using \eqref{eqn:solutionLinear} we can explicitly write the expression of kinetic component of the half chain as a function of time:
\begin{equation}\label{eqn:halfChainEv}
    K^\text{os}_\text{half}(t) = T \sum_{j=1}^{N/2} \sum_{k,k'=1}^N C_{jk} C_{jk'} \sin(\omega_k t -\pi/2)  \sin(\omega_{k'} t -\pi/2), 
\end{equation}
Thus, we verify that $K^\text{os}_\text{half}(0)$ is macroscopically different from the thermalized value in \eqref{eqn:theoPredictionHalfChain}, and the same goes for $U^\text{os}_\text{half}(0)$. The approach to thermalization can then be thought of as a dephasing process that removes the time dependence and introduces a factor $\delta_{k k'}/2$ in \eqref{eqn:halfChainEv}.

\begin{figure}
    \centering
    \includegraphics[scale=0.5]{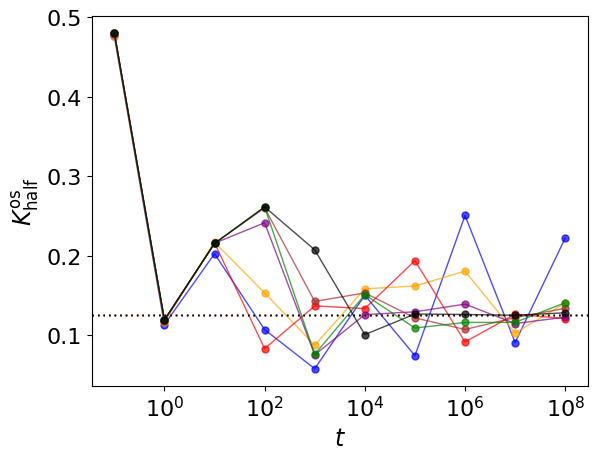}
    \includegraphics[scale=0.5]{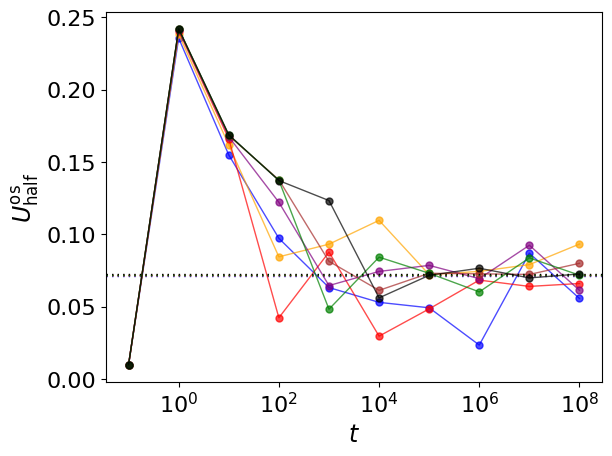}
    \includegraphics[scale=0.5]{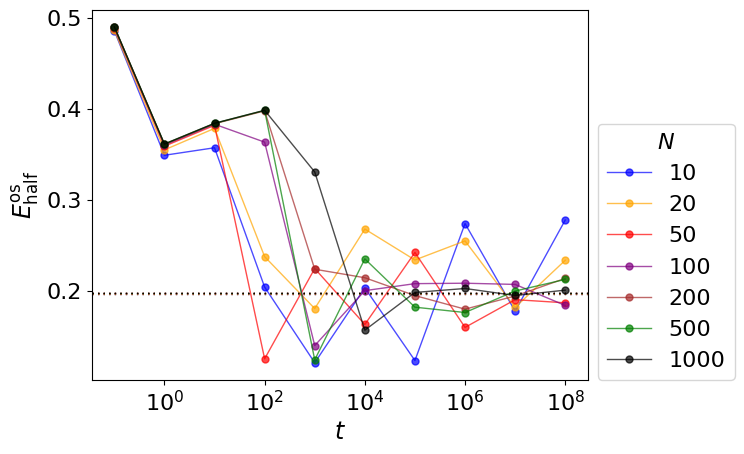}
    \caption{Solid lines: evolution of the on-site energy of the half-chain $E_\text{half}^\text{os}$ and of its kinetic $K_\text{half}^\text{os}$ and potential $U_\text{half}^\text{os}$ contributions over time, for varying $N$. The initial conditions are set according to \eqref{eqn:initialConditionLinearCase} with $T=0.5/N$ and $k_R=2$. Dotted lines: theoretical prediction for the long-time average according to \eqref{eqn:theoPredictionHalfChain} and \eqref{eqn:half_energy}. The observables approach thermalization at their expected microcanonical value for $N\gg 1$.}
    \label{fig:linearCase}
\end{figure}

We illustrate the complete thermalization process in Fig.~\ref{fig:linearCase}, where we plot separately the evolution of $K^\text{os}_\text{half}(t)$, $U^\text{os}_\text{half}(t)$, and $E^\text{os}_\text{half}(t)$ for growing $N$. 
We want to keep the total energy $E$ fixed, so we assume that $T$ scales as $1/N$. We observe that thermalization is approached and fluctuations vanish for $N\gg 1$ as time increases.  As a final remark, different out-of-equilibrium initial conditions may be explored by changing the sign of the $P_j(0)$'s in \eqref{eqn:initialConditionLinearCase}.

\subsubsection{Towards energy equipartition from a localized initial excitation}
\label{sec:harm_loc}

Let us now focus on a class of initial conditions for which the long-time averages of the on-site energies do not approach their microcanonical equilibrium values. In other words, we pick some initial conditions that belong to a subset of the microcanonical energy shell whose measure goes to zero for $N\rightarrow\infty$, according to \eqref{eqn:nontypical}.%\eqref{eqn:KhinchinOnSiteEnergies}.

Specifically, we choose the initial condition for which all particles are initially at rest, and only the potential energy of the $r$th particle is non-zero: 
\begin{equation}\label{eqn:initialConditionEnEqui}
    q_j(0)=\sqrt{\frac{2E}{c_\mathcal{N}}}\delta_{jr},\quad p_j(0)=0 \text{ for all }j, 
\end{equation}
where $$c_\mathcal{N}=\sum_k C_{rk}^2\omega_k^2$$ is a normalization constant for the total energy in the system.
Therefore,
\begin{equation}
\label{eqn:initialConditionEnEquiNM}
    Q_j^2(0)=\frac{2NT}{c_\mathcal{N}} C_{rj}^2,\quad P^2_j(0)=0 \text{ for all }j. 
\end{equation}

Next, using \eqref{eqn:longTimeEnergyOS} we compute the long-time average of the on-site energies for this initial condition (we refer the readers to Appendix~\ref{sec:crossProductProof} for the derivation): 
\begin{equation}\label{eqn:long_time_energy_equipartition}
    \overline{E_j^\text{os}}_\infty = NT\sum_{k=1}^N \frac{C_{jk}^2 C_{rk}^2}{2 c_\mathcal{N}}(\omega_k^2+k_R) = \begin{cases}
         \dfrac{3\,T\,N\,(1+k_R)}{2(2+k_R)(N+1)}\xrightarrow{N \to \infty} \dfrac{3\,T\,(1+k_R)}{2\,(2+k_R)} &\text{ if }j=r \text{ or }j+r=N+1,\\[0.4cm]
            \dfrac{T\,N\,(1+k_R)}{(2+k_R)(N+1)}\xrightarrow{N \to \infty} \dfrac{T\,(1+k_R)}{(2+k_R)}& \text{ otherwise}.
    \end{cases} 
\end{equation}

From the above expression, we observe that almost all on-site energies equilibrate at the value $\frac{T(1+k_R)}{(2+k_R)}$ in the thermodynamic limit. Two on-site energies equilibrate at a slightly different value, which anyway scales in the same way in the thermodynamic limit. 

We now focus on the \textit{renormalized effective number of degrees of freedom} \cite{Baldovin2021}, defined as
\begin{equation}
    \label{eqn:neff}
    n_\text{eff}(t)=\frac{\exp\left(-\sum_k u_k(t) \log u_k(t) \right)}{N},
\end{equation}
with 
\begin{equation}
    \label{eqn:uk}
    u_k(t) = \frac{\overline{E_j^\text{os}}_t}{\sum_{j=1}^N\overline{E_j^\text{os}}_t}.
\end{equation}
From \eqref{eqn:canonicalOnSiteEnergies} we can show\footnote{This follows from the fact that the variance of the canonical expectation value for the on-site energies over $j=1,\ldots,N$  divided by the square of the mean value goes to zero for $N\rightarrow\infty$.} that in the canonical ensemble $\langle n_\text{eff}\rangle_\text{can}\rightarrow1$ for $N\rightarrow\infty$. From \eqref{eqn:long_time_energy_equipartition}, $n_\text{eff}(\infty)$ also tends to 1 for $N\gg 1$, if we excite only a single particle at time $t=0$ according to \eqref{eqn:initialConditionEnEqui}. Therefore, for the initial condition in \eqref{eqn:initialConditionEnEqui}, $n_\text{eff}$ starts from a value that is extremely out of equilibrium ($n_\text{eff}(0)\rightarrow 0$ for $N\gg 1$), and then thermalizes to the canonical value $n_\text{eff}=1$ over a long time. 

It is worth remarking that the dynamics we are studying is linear and separable, so chaos does not play a role in the emergence of thermalization for $n_\text{eff}$. This said, different observables do not display thermalization at long times. For instance, the on-site energies of the normal modes do not reach equipartition at the same value predicted by the canonical ensemble in \eqref{eqn:canonicalOnSiteEnergies}. In this case, for $N\gg 1$, we observe energy equipartition (excluding a couple of sites that become negligible in the thermodynamic limit) without thermalization. 

In \cite{Cattaneo2025thermalization}, we obtained similar results for both classical and quantum harmonic systems. In particular, in \cite{Cattaneo2025thermalization} we considered initial conditions such that the ``populations'' of the normal modes, i.e. the coefficients $A_k^2$ in \eqref{eqn:longTimeEnergyOS}, are roughly equal. By this we mean populations that do not depend significantly on the mode $k$: the relative difference between different $A_k^2$ is of order $\mathcal{O}(1)$ and does not depend on $N$. For these initial conditions, we showed that in the long-time limit the on-site energies reach equipartition without thermalization. The initial conditions for $Q_j^2(0)$ in \eqref{eqn:initialConditionEnEqui} fall into this class.

\subsection{Nonlinear case}
Let us now consider the modified FPUT model~\eqref{eqn:HamFPUT} in the presence of quartic terms, namely $H_{0,\beta}$ with $\beta>0$. In what follows, we consider the case of small nonlinearity, i.e. $\beta \ll 1$. Rescaling positions and momenta shows that this limit is equivalent to having $\beta$ of order $O(1)$ and a small specific energy $\varepsilon=E/N$. Although the latter choice is often preferred when studying the original FPUT problem, here we treat $\beta$ as the small parameter. We do so because this choice is more adapted to a perturbative expansion around the harmonic limit. In this section, we ask whether the inclusion of small nonharmonic terms in the Hamiltonian is sufficient to ensure thermalization (and on what time scales).

A result that we need in the following is the expression for the average value of the observable $E_j^\text{os}$ defined by Eq.~\eqref{eqn:energyOS}, and evaluated in the microcanonical ensemble:
\begin{align}
\label{eq:eos_beta}
    %\av{E^\text{os}_j}_{\beta}=\frac{E}{2N}\cbr{1+k_R \mathsf{M}^{-1}_{jj}}+\frac{\beta}{4}\cbr{\frac{E}{N}}^2\frac{1}{N+1}\sbr{N\cbr{\av{H_4}_\text{can}\av{q_j^2}_\text{can}- \av{H_4 q_j^2}_\text{can}}+\cbr{1+k_R\mathsf{M}^{-1}_{jj}}\av{H_4}_\text{can} } + O(\beta^2)\,,
    \av{E^\text{os}_j}_{\beta}&=\frac{E}{2 N}\left(1+k_R \mathsf{M}^{-1}_{jj}\right)
    \nonumber\\
 &   +\frac{\beta N}{2(N+1)E}\left[k_RN\left(\langle q_j^{2}\rangle_\text{can}\langle  H_4(\mathbf{q})\rangle_\text{can}-\langle q_j^{2}  H_4(\mathbf{q})\rangle_\text{can}\right)+\frac{E}{N}(1+k_R \mathsf{M}^{-1}_{jj})\langle  H_4(\mathbf{q})\rangle_\text{can}\right]+O(\beta^2)\,,
\end{align}
where
\begin{equation}
    H_4(q)=\frac{1}{4}\sum_{j=0}^N\cbr{q_j-q_{j+1}}^4\,.
\end{equation}
We use again the symbol $\av{\cdots}_{\text{can}}$ to refer to canonical averages \textit{in the unperturbed harmonic limit}, while we denote by $\av{\cdots}_{\beta}$ microcanonical averages for generic $\beta$.
We derive the perturbative relation~\eqref{eq:eos_beta} in Appendix~\ref{sec:app_pert}. The $\beta = 0$ case is the microcanonical average in the harmonic limit, which coincides with the canonical average in Eq.~\eqref{eqn:canonicalOnSiteEnergies} .

\subsubsection{Thermalization}

\begin{figure}[h!]
    \centering
    \includegraphics[width=0.7\linewidth]{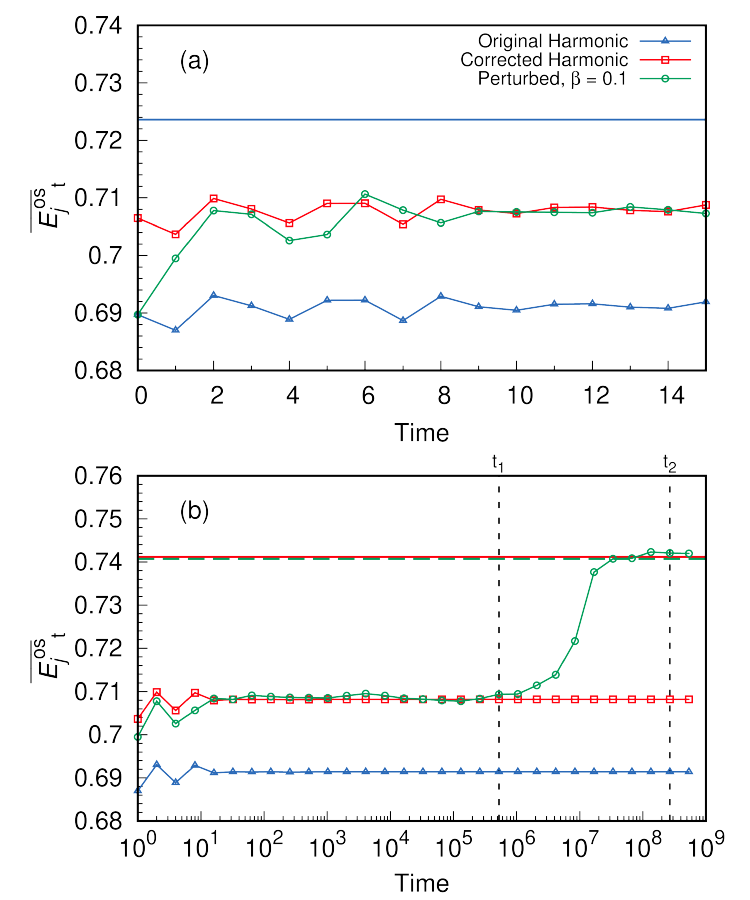}
    \caption{Time average of the observable $E^{\text{os}}_j$ defined in Eq.~\eqref{eqn:energyOS}, as a function of time $t$, averaged on all sites $j$. Averages are estimated as sample means over integer times. Panel (a) focuses on short time scales, panel (b) on the long-time behaviour. In both panels, blue triangles represent the unperturbed harmonic dynamics (energy $E_0$), while green circles stand for the perturbed non-harmonic one. The harmonic dynamics with corrected total energy $E=E_0+\Delta$ is represented by red squares. Horizontal lines indicate the corresponding analytical value of \mar{$\av{E_k^{\text{os}}}_\beta$}, provided by Eq.~\eqref{eq:eos_beta}, with color code as before. The two times $t_1$ and $t_2$ highlighted  in panel (b) are better explored in Fig.~\ref{fig:modes}. Parameters: $\beta=0.1$, $E_0/N=1$, $N=256$.}
    \label{fig:observable}
\end{figure}

We have shown in Sec.~\ref{sec:harm_loc} that if all energy is initially provided to a single particle (localized initial excitation), in the harmonic system with $\beta=0$, the observable $E^{\text{os}}$ defined by Eq.~\eqref{eqn:energyOS} does not thermalize: 
$$\overline{E_j^\text{os}}_{\infty} \ne \av{E^{\text{os}}}_\text{can}.$$
We expect the same to occur whenever the initial distribution of the energy among the normal modes is highly non uniform. When the distribution is close to uniform, within some degree of tolerance, we expect instead  a typical behaviour indistinguishable from equilibrium~\cite{Cattaneo2025thermalization}.
Let us note that these initial conditions %used in the  numerical computations showed in Fig.~\ref{fig:observable} 
are very different
from those used in the original works by Fermi et al.~\cite{FPU55} where only few normal modes  at low frequency are
initially excited.

Now, let us assume that at a certain time of the evolution (e.g., $t=0$) we can switch on the perturbation, and then let the system evolve for a time interval of duration $t^{\star}$, following the new, weakly nonlinear dynamics. At time $t^{\star}$ the perturbation is switched off again, and one verifies whether  the observable $E_j^{\text{os}}$ has thermalized.  The inclusion of nonlinear contributions makes the system non integrable: it is therefore reasonable to expect that, for $t^{\star}$ long enough, the above protocol guarantees thermalization. The transient behaviour is however not obvious. For $\beta \ll 1$, one would expect  the dynamics  to stay ``close'' to the unperturbed one for a certain amount of time, at least until none of the $\beta (q_{j}-q_{j+1})^4/4$ terms in the Hamiltonian is comparable with the specific energy. Eventually, the system will  reach a configuration where the effect of some of the quartic terms is relevant, and the difference with the unperturbed Hamiltonian becomes manifest. This should allow the dynamics to escape from the vicinity of the unperturbed torus and explore the whole phase space hypersurface at constant energy.

To test the above qualitative scenario, we perform numerical simulations of a system of coupled oscillators evolving according to the Hamiltonian~\eqref{eqn:HamFPUT}.
%Inspired by the case of localized initial excitation, 
We choose the out-of-equilibrium initial condition in such a way that the energies of the normal modes are given by
\begin{equation}\label{eq:initialCondFPUTexp}
    E^{\text{nm}}_k(0)=\frac{\omega_k E_0}{\sum_j \omega_j}\,,
\end{equation}
where $E_0$ is the total energy of the harmonic system. The initial phase of each normal mode is extracted randomly. The above condition implies that the action variable
\begin{equation}
    \mathcal{I}^{\text{nm}}_k= \oint d Q_k\, P_k=2 \pi \frac{E^{\text{nm}}_k}{\omega_k}\,,
\end{equation}
defined as the contour integral along the periodic orbit of the $k$-th mode, is independent of $k$.
Qualitatively, this situation resembles the one in Eq.~\eqref{eqn:initialConditionEnEquiNM}, where the fast modes are more excited than the slow ones: as such, it can be seen as ``opposite'' to the one in the FPUT experiment. We then evolve the dynamics via a Velocity Verlet Update, choosing the time step in such a way that the total energy of the system is conserved with a relative error of the order of $10^{-5}$.
%The initial distribution of the energies among the normal modes is consistent with Eq.~\eqref{eqn:initialConditionEnEquiNM}, if one makes the approximation that all $C_{rj}^2$ elements are equal. In this sense, it mimics the effect of an initial localization.

%\iffalse
%The initial condition in \eqref{eq:initialCondFPUTexp} corresponds to the amplitudes 
%\begin{equation}
%    A_k^2 = \frac{2 E}{\omega_k\sum_j \omega_j}, \text{ for all }k.
%\end{equation}
%Then, from \eqref{eqn:longTimeEnergyOS} we can immediately find (see Appendix~\ref{sec:crossProductProof} for details) the predicted long-time average of the on-site energies:
%\begin{equation}
%    \label{eqn:longTimeAvFPUT}
%    \overline{E_k^\text{os}}_\infty = T \frac{k_R+1}{k_R+2},\quad T=\frac{E}{N}.
%\end{equation}
%As the above expression does not depend on $k$, the on-%site energies are exactly equipartitioned. 
%\fi

 We compute the observable $\overline{E_j^\text{os}}_{t}$ (averaged over all sites $j$) at different waiting times $t$. The original, unperturbed harmonic dynamics immediately relaxes to a stationary value, which, as expected, is different from the ensemble average [Fig.~\ref{fig:observable}(a)]. At time $t=0$, the nonlinear perturbation is switched on: the perturbed curve quickly reaches a plateau at a value of  $\overline{E_j^\text{os}}_{t}$ different from the $t\to \infty$ limit provided by Eq.~\eqref{eq:eos_beta}. 

When the nonlinearity is switched on, the total energy of the system changes from $E_0$ to $E=E_0+\Delta$, as an effect of the sudden inclusion of the quartic perturbative corrections. In Fig.~\ref{fig:observable} we compare therefore the evolution of the perturbed nonlinear dynamics with a harmonic system whose total energy is $E$. We see that the transient plateau that the nonlinear dynamics reaches is basically given by the value of $\overline{E_j^\text{os}}_{t}$ in such harmonic chain with corrected energy, corresponding to equipartition among the on-site energies.

In the long time limit, the ensemble average~\eqref{eq:eos_beta} is eventually reached [Fig.~\ref{fig:observable}(b)]. Again, the total energy $E$ to consider in the microcanonical average is the one that takes into account the nonharmonic corrections.

\begin{figure}
    \centering    \includegraphics[width=0.99\linewidth]{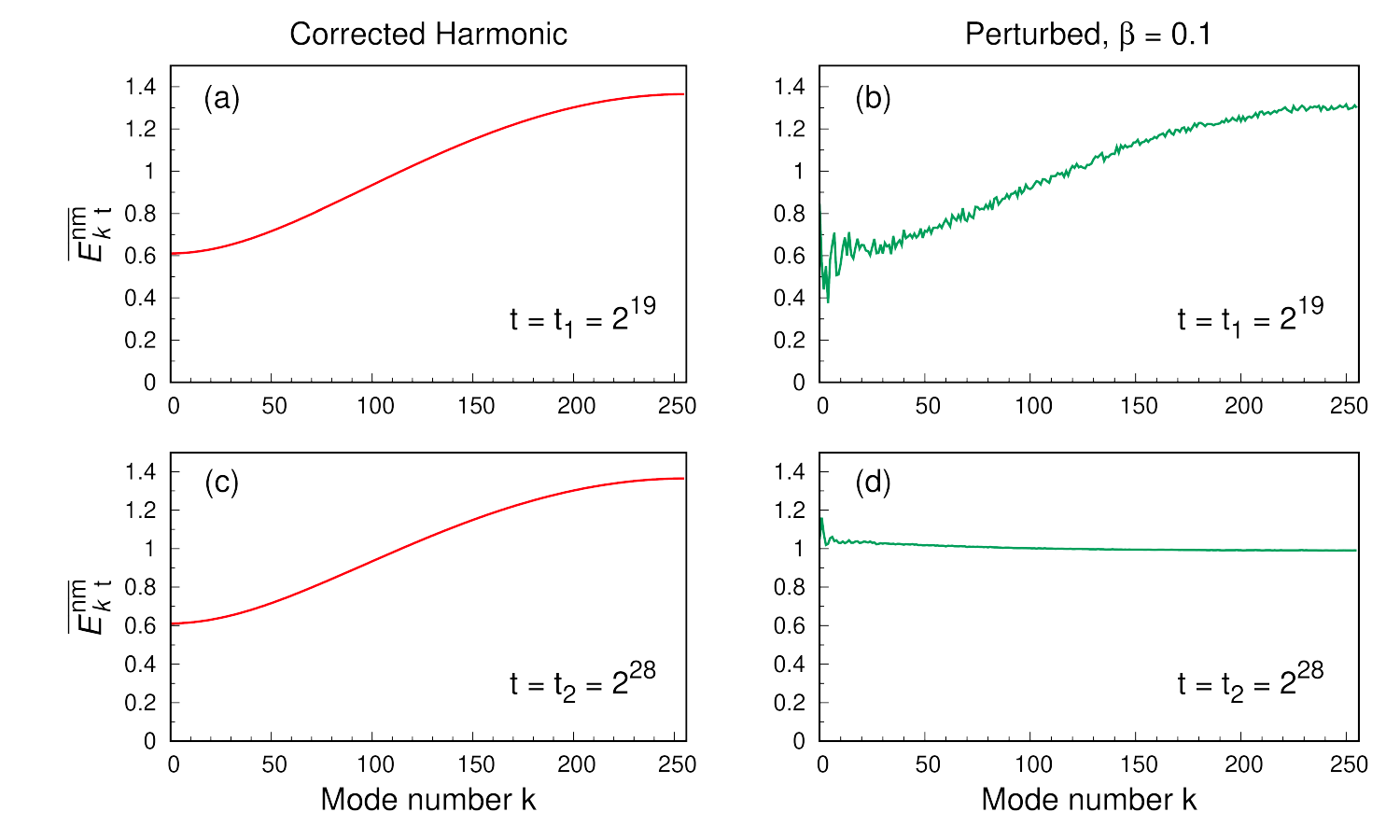}
    \caption{Time-averaged distribution of the harmonic energy among the normal modes. Both the harmonic case with corrected energy [panels (a) and (c)] and the nonlinearly perturbed evolution [panels (b) and (d)] are considered. We focus on the two times $t_1=2^{19}$ [panels (a-b)] and $t_2=2^{28}$ [panels (c-d)] highlighted in Fig.~\ref{fig:observable}. Parameters as in Fig.~\ref{fig:observable}.}
    \label{fig:modes}
\end{figure}

To further delve into the thermalization dynamics in the presence of nonlinear corrections, in Fig.~\ref{fig:modes} we show the distribution of the energy among the normal modes for two selected waiting times, $t_1$ and $t_2$: the former is just before the perturbed dynamics departs from the harmonic behaviour, the latter is chosen after reaching the thermal state. While at time $t_1$ the distribution is close to the unperturbed case, at time $t_2$ almost perfect equipartition between the modes has been reached. The results shown in Fig.~\ref{fig:modes} also provide an indirect check that the observed effect (departure from the non-equilibrium metastable phase) is not numerical: indeed, the harmonic simulation maintains the initial excitation of the modes for extremely long times, as expected.

\iffalse
The system has been largely studied, and the overall scenario can be summarized as follows.
Detailed numerical computations suggest  that
the behaviour  of the FPUT  system, with a fixed strength of the non quadratic
term $\beta$ in the Hamiltonian, for a given number of particles 
$N$ and  energy density $\epsilon= E/N$, can be summarized  as follow:
there is a threshold  $\epsilon_t$  such that 
I)   if $\epsilon < \epsilon_t$  KAM tori play a major role and the system does can reach equipartition
only   after a very long time $\tau_R$;
II)  if $\epsilon < \epsilon_t$  KAM tori have a minor  effect and the system follows equipartition 
after a short time
\fi

An important technical  aspect concerns  the
relaxation time $\tau_R$ to reach equipartition of the normal modes. It is known since the FPUT numerical experiment that, when initializing the system far from equilibrium, the relaxation time can be very long, and 
might also depend on the details of the initial condition~\cite{Gallavotti2008,kantz1994equipartition,de1999finite,ruffo2001time}.
This is a consequence of the fact that   Hamiltonian systems do not
have attractors: therefore, the way the system is prepared (particularly for large $N$) 
can have nontrivial consequences on the dynamics.
For the classical FPUT initial conditions, it has been shown through careful numerical experiments that if the initially excited normal modes are always between
$k_1$ and $k_2=k_1+ \Delta k$,  one has
$$
\tau_R \propto N^a \cbr{\frac{E}{N}}^{-b}
$$
where $a$ and $b$ are related to the dependence of $k_1$ and $\Delta k$ on $N$. If $k_1$  and $k_2$ are fixed, one has $a= 1/2$ and $b=1$~\cite{kantz1994equipartition,de1999finite,ruffo2001time}. Crucially, the dependence on $N$ is only power-law.

\begin{figure}
    \centering    \includegraphics[width=0.6\linewidth]{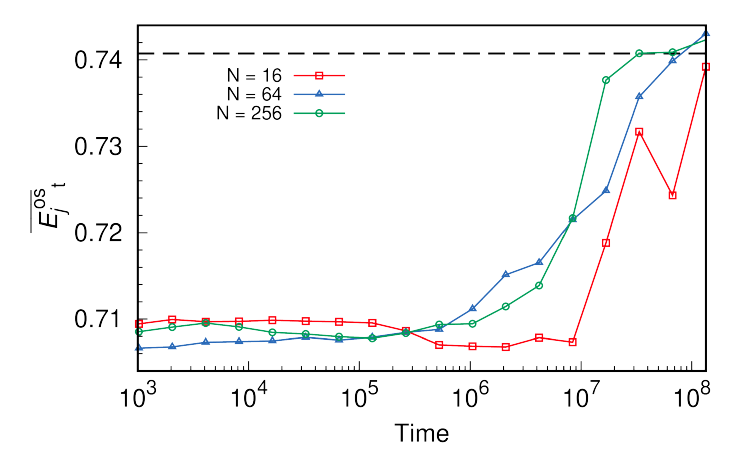}
    \caption{Dependence of thermalization time on $N$. The simulation already shown in Fig.~\ref{fig:observable} is compared with two analogous cases with same parameters but different values of $N$. For single realizations, no scaling of the thermalization time with $N$ is appreciable.}
    \label{fig:observableN}
\end{figure}

In Fig.~\ref{fig:observableN} we explore the dependence on $N$ of the relaxation time, by repeating our simulation with chains of different lengths. Our results do not allow to infer any particular scaling with $N$. We can, however, exclude an exponential dependence, as expected in analogy with the classical FPUT initial conditions. We leave a more systematic study of this point for future work.

\section{Conclusions}\label{sec:conclusion}
We have provided some examples that, in our opinion, help to clarify the reasons for the success of SM in predicting the features of  macroscopic systems.
In particular, starting from well established results for non equilibrium systems, we have investigated the role of chaos in thermalization properties of systems with $N\gg 1$ particles.
First, we have carefully analized the thermalization properties of a chain of harmonic oscillators, taking into account different initial conditions which are far from equilibrium, for relevant macroscopic observables such as the kinetic energy or the self-energy of the particles (i.e., the contributions to energy which do not involve interactions).

Then, we have focused on a specific class of initial conditions, inspired by the spectral energy distribution of a fully localized initial condition, and we have examined the effect of nonlinearities (i.e., of chaos).
Differently from the seminal contribution of Fermi-Pasta-Ulam-Tsingou~\cite{FPU55}, we have considered initial conditions in which every normal mode $k$ has an energy proportional to its frequency $\omega_k$.  
%By means of analytical calculation, we showed that this set of initial conditions behaves similarly to FPUT in which only the $j-$th physical particle is initially excited. More specifically, the two set of initial conditions have the same thermalization property: some observables, e.g. the effective number of degrees of freedom $n_\text{eff}$,  reach the values predicted by SM, while some others, e.g. the local energy $E^\text{os}_j$, do not. 
We have performed numerical computations on a modified FPUT chain in a regime in which nonlinear interactions are small compared to the harmonic ones. 
In such a case, our results show that the system remains close to the unperturbed linear dynamics for a long period of time, during which the time averages of observables equal those computed in the harmonic system (similarly to pre-thermalization discussed in Quantum SM~\cite{Mori2018}). At very large times, nonlinear terms become non-negligible (as expected also from numerical results in~\cite{Benettin2011}) and the temporal averages converge to the values predicted by the microcanonical ensemble. 

%It is worth mentioning that although invoking choaticity is appealing as it guarantees good thermalization properties for \textit{any} observables, it is a very strong requirement for a physical dynamics. On one hand, determining whether a given system is mixing is a rather difficult task. On the other hand, it leads also systems with a small number of degrees of freedom, for which a thermodynamics description is not properly defined~\cite{}, to thermalized state.

Our results show that, even in chaotic systems, the timescale for reaching the equilibrium state might be much larger than the observation time, and the long-lasting metastable state can be, in this sense, very misleading. For these reasons, we argue that chaoticity is not the fundamental ingredient for the irreversible relaxation observed in physical systems. Instead, 
according to Khinchin's point of view, that validity of the   predictions of the SM seems to rely on
 the large $N$  limit and on the choice of physically
meaningful observables. 
The details of 
the microscopic dynamics,  namely chaos and ergodic property,
appear to be a highly restrictive requirement, and they do not seem to play the main role.

%These results support a generalization of  Khinchin's  approach to cases with far from equilibrium initial conditions. 

\appendix

\section{Analytical derivations}
\subsection{Derivation of Eq.~\eqref{eqn:IPR_1dChain_FBC}}\label{sec:IPR_proof}
We compute 
\begin{equation}
    \begin{split}
        \text{IPR}_j =& \sum_{k=1}^N C_{jk}^4 = \frac{1}{2(N+1)^2}\left[3N+\sum_{k=1}^N\cos\left(\frac{4\pi kj}{N+1}\right) -4 \sum_{k=1}^N\cos\left(\frac{2\pi kj}{N+1}\right)\right], 
    \end{split}
\end{equation}
where we have used $\sin^4x=(3+\cos 4x -4\cos 2x)/8$. Next, 
\begin{equation}\label{eqn:trickProof}
\begin{split}
    \sum_{k=1}^N e^{i \frac{2\pi  k  \alpha}{N+1}}=&
    \sum_{k=1}^N (e^{i \frac{2\pi \alpha}{N+1}})^k = \frac{1-e^{i 2\pi \alpha }}{1-e^{i(2\pi \alpha)/(N+1)}}-1 =-1\quad\text{ for any integer }0<\alpha<N+1.   
\end{split}
\end{equation}
From the above equations we immediately recover \eqref{eqn:IPR_1dChain_FBC}.

\subsection{Derivation of Eq.~\eqref{eqn:long_time_energy_equipartition}}
\label{sec:crossProductProof}
We compute 
\begin{equation}
    \begin{split}
        \sum_{k=1}^N C_{jk}^2 C_{j'k}^2 = &\sum_{k=1}^N\frac{4}{(N+1)^2}\sin^2\left(\frac{\pi kj}{N+1}\right)\sin^2\left(\frac{\pi kj'}{N+1}\right)\\
        =&\sum_{k=1}^N\frac{1}{(N+1)^2}\left(1-\cos\frac{2\pi j k}{N+1}-\cos\frac{2\pi j' k}{N+1}+\cos\frac{2\pi j k}{N+1}\cos\frac{2\pi j' k}{N+1}\right).
    \end{split}
\end{equation}
Next, we observe
\begin{equation}
    \sum_{k=1}^N \cos\frac{2\pi j k}{N+1}\cos\frac{2\pi j' k}{N+1}=\frac{\Re[A+B]}{2},
\end{equation} where 
\begin{equation}
    A = \sum_k e^{i \frac{2\pi k (j+j')}{N+1}},\quad
    B = \sum_k e^{i \frac{2\pi k (j-j')}{N+1}}.
\end{equation}
Using \eqref{eqn:trickProof}, 
\begin{equation}
    A = \begin{cases}
        N & \text{ if }j+j'=N+1,\\
        -1 & \text{ otherwise},
    \end{cases}\qquad  B = \begin{cases}
    N& \text{ if }j=j',\\
    -1 & \text{ otherwise}.
      \end{cases}\
\end{equation}
So, 
\begin{equation}
        \sum_{k=1}^N C_{jk}^2 C_{j'k}^2 = \begin{cases}
            \frac{3}{2(N+1)} &\text{ if }j=j' \text{ or }j+j'=N+1,\\
            \frac{1}{N+1} & \text{ otherwise}.
        \end{cases}
\end{equation}

It can be shown that a similar expression arises when we include also the frequencies of the normal modes according to \eqref{eqn:longTimeEnergyOS}:
\begin{equation}\label{eqn:derivationWithFreq}
    \begin{split}
        \sum_{k=1}^N C_{jk}^2 C_{j'k}^2 (\omega_k^2-k_R) = &\sum_{k=1}^N\frac{16}{(N+1)^2}\sin^2\left(\frac{\pi kj}{N+1}\right)\sin^2\left(\frac{\pi kj'}{N+1}\right)\sin^2\left(\frac{\pi k}{2(N+1)}\right)\\
        =&\sum_{k=1}^N\frac{2}{(N+1)^2}\left(1-\cos\frac{2\pi j k}{N+1}-\cos\frac{2\pi j' k}{N+1}-\cos\frac{\pi k}{N+1}+\cos\frac{2\pi j k}{N+1}\cos\frac{2\pi j' k}{N+1}\right.\\
        &\left.+\cos\frac{2\pi j k}{N+1}\cos\frac{\pi  k}{N+1}+\cos\frac{2\pi j k}{N+1}\cos\frac{\pi  k}{N+1}-\cos\frac{2\pi j k}{N+1}\cos\frac{2\pi j' k}{N+1}\cos\frac{\pi k}{N+1}\right)\\
        =&\begin{cases}
            \frac{3}{N+1} &\text{ if }j=j' \text{ or }j+j'=N+1,\\
            \frac{2}{N+1} & \text{ otherwise}.
        \end{cases}
    \end{split}
\end{equation}
The last expression can be obtained by noticing that $\cos(\pi k/(N+1))=-\cos(\pi (N+1-k)/(N+1))$, so all summations including the term $\cos(\pi k/(N+1))$ in \eqref{eqn:derivationWithFreq} are equal to zero.

Finally, a similar calculation immediately yields
\begin{equation}
    \sum_{j=1}^N\omega_j^2 = N(k_R+2),\quad c_\mathcal{N} = \sum_{j=1}^N C_{rj}^2\omega_j^2 = 2+k_R,\quad \text{ for all }N,\text{ for all }r.
\end{equation}

\subsection{Microcanonical estimates: derivation of Eq.~\eqref{eq:eos_beta}}
\label{sec:app_pert}

\subsubsection{Unperturbed limit}
The microcanonical averages can be computed using the formalism of Laplace transform. Before computing the expected value of our observable in the non-linear dynamics, it is convenient to define the microcanonical average of the linear case. 
The microcanonical partition function $Z^{(m)}_0(E)$ of the linear system is
\begin{subequations}
    \begin{align}
    Z_0(E)&=\int{\rm d}\mathbf{q}{\rm d}\mathbf{p}\, \delta\left(E-H_0(\mathbf{q},\mathbf{p})\right)=\label{eq:Z0-def-delta}\\
    &=\int_{\gamma-i \infty}^{\gamma+i \infty} {\rm d}y\,\frac{e^{yE}}{2\pi i}\int{\rm d}\mathbf{q}{\rm d}\mathbf{p}\, e^{-yH_0(\mathbf{q},\mathbf{p})}=\label{eq:Z0-def}\\
    &=\int_{\gamma-i \infty}^{\gamma+i \infty} {\rm d}y\,\frac{e^{yE}}{2\pi i}\,y^{-N}\int{\rm d}\tilde{\mathbf{q}}{\rm d}\tilde{\mathbf{p}}\, e^{-H_0(\tilde{\mathbf{q}},\tilde{\mathbf{p}})}=\label{eq:Z0-change-variable}\\
    &=\frac{(2\pi)^N}{ \sqrt{\det(K)\det(M)}}\frac{E^{N-1}}{\Gamma(N)}\label{eq:Z0-final}
\end{align}
\end{subequations}
where $\gamma$ is a real positive constant. Note that for passing from Eq.~\eqref{eq:Z0-def} to Eq.~\eqref{eq:Z0-change-variable} we have performed the change of variables $\tilde{\mathbf{q}}=\sqrt{y}\mathbf{q}$ and $\tilde{\mathbf{p}}=\sqrt{y}\mathbf{q}$. To pass from Eq.~\eqref{eq:Z0-change-variable} to Eq.~\eqref{eq:Z0-final} we have used the fact that the definition of the Euler $\Gamma$ function can be seen as the Laplace transform of a power: taking its antitransform one has
\begin{equation}
    \int_0^{\infty} E^{N-1}e^{-yE}dE=\frac{\Gamma(N)}{y^{N}}\iff \int_{\gamma-i \infty}^{\gamma+i \infty} {\rm d}y\,\frac{e^{yE}}{2\pi i}\frac{\Gamma(N)}{y^{N}}=E^{N-1}\,.
    \label{eq:inverse_laplace}
\end{equation}
To simplify the following calculations, 
let us define $O\equiv O_{2n}(\mathbf{q},\mathbf{p})$  as a generic linear combination of monomials of degree $2n$ in $\mathbf{q}$ and $\mathbf{p}$, i.e.
\begin{align}
&O_{2n}(\mathbf{q},\mathbf{p})=\sum_{\substack{ \mathbf{m}\\ S(\mathbf{m})=2n}}c\,_{\mathbf{m}}\, q_1^{m_{1}} p_1^{m_{2}} \cdots q_N^{m_{2N-1}} p_N^{m_{2N}}\,.
%&\text{where } l+m=\sum_{j=1}^N (l_j+m_j)\nonumber\,.
\end{align}
with $S(\mathbf{m})=\sum_{j=1}^N (m_j)$, for some choice of the coefficients $c\,_{\mathbf{m}}$.
Moreover, let us define the functional $\mathcal{F}^n_m[O]$ as
\begin{subequations}
\begin{align}
    \mathcal{F}^{n}_m[O]&=\frac{1}{Z_0(E)}\int_{\gamma-i \infty}^{\gamma+i \infty} {\rm d}y\,\frac{e^{yE}}{2\pi i}\,y^{m}\int{\rm d}\mathbf{q}{\rm d}\mathbf{p}\, e^{-yH_0(\mathbf{q},\mathbf{p})}O_{2n}(\mathbf{q},\mathbf{p})=\label{eq:def_functional}\\
    &=\frac{1}{Z_0(E)}\left(\frac{N}{E}\right)^{(N+n)}\int_{\gamma-i \infty}^{\gamma+i \infty} {\rm d}y\,\frac{e^{yE}}{2\pi i}\,y^{-(N+n-m)}\int{\rm d}\hat{\mathbf{q}}{\rm d}\hat{\mathbf{p}}\, e^{-\frac{N}{E}H_0(\hat{\mathbf{q}},\hat{\mathbf{p}})}O_{2n}(\hat{\mathbf{q}},\hat{\mathbf{p}})=\label{eq:def_functional_change_variable}\\
    &=\frac{\Gamma(N) \sqrt{\det(K)\det(M)} }{E^{N-1} (2\pi)^N}\left(\frac{N}{E}\right)^{(N+n)}\frac{E^{(N+n-m-1)}}{\Gamma(N+n-m)}\int{\rm d}\hat{\mathbf{q}}{\rm d}\hat{\mathbf{p}}\, e^{-\frac{N}{E}H_0(\hat{\mathbf{q}},\hat{\mathbf{p}})}O_{2n}(\hat{\mathbf{q}},\hat{\mathbf{p}})=\label{eq:def_functional_explicit}\\
    &=\frac{N^n}{E^m}\frac{\Gamma(N)}{\Gamma(N+n-m)}\left[\left(\frac{N}{2\pi E}\right)^{N}\sqrt{\det(K)\det(M)}\int{\rm d}\hat{\mathbf{q}}{\rm d}\hat{\mathbf{p}}\, e^{-\frac{N}{E}H_0(\hat{\mathbf{q}},\hat{\mathbf{p}})}O_{2n}(\hat{\mathbf{q}},\hat{\mathbf{p}})\right]=\\
    &=\left(\frac{N}{E}\right)^{m}\frac{N^{n-m} \Gamma(N)}{\Gamma(N+n-m)}\langle O_{2n}\rangle_\text{can}\label{eq:def_functional_final}
\end{align}
\end{subequations}
where from Eq.~\eqref{eq:def_functional} to Eq.~\eqref{eq:def_functional_change_variable} we have performed the change of variables $\hat{\mathbf{q}}=\sqrt{\frac{Ey}{N}}\mathbf{q} $ and $\hat{\mathbf{p}}=\sqrt{\frac{Ey}{N}}\mathbf{q}$, from Eq.~\eqref{eq:def_functional_change_variable} to Eq.~\eqref{eq:def_functional_explicit} we used Eq.~\eqref{eq:inverse_laplace}, and the canonical average coincides with those introduced in~\ref{sec:FPUT}. Here and in what follows, we use the compact notation $\av{\cdots}_c=\av{\cdots}_{\text{can}}$.
By defining $O_p\equiv p_j^{2n}$ and $O_q\equiv q_j^{2n}$, 
%Using the definitions above and denoting  $O_p\equiv p_j^{2n}$, 
the microcanonical averages of $\langle p_j^{2n}\rangle$ and $\langle q_j^{2n}\rangle$ can be computed as follows
\begin{align}
    &\langle p_j^{2n}\rangle=\mathcal{F}^{n}_0[O_p]=\frac{N^n\Gamma(N)}{\Gamma(N+n)}\langle p_j^{2n}\rangle_{c}\,;\\
    &\langle q_j^{2n}\rangle =\mathcal{F}^{n}_0[O_q]=\frac{N^n \Gamma(N)}{\Gamma(N+n)}\langle q_j^{2n}\rangle_{c}\,.
\end{align}
\begin{comment}
\begin{align}
    \langle p_j^{2n}\rangle %&=\frac{1}{Z_0(E)}\int {\rm d}y\,\frac{e^{yE}}{2\pi i}\int{\rm d}\mathbf{q}{\rm d}\mathbf{p}\, e^{-yH_0(\mathbf{q},\mathbf{p})}\,p_j^{2n}=\\
    &=\mathcal{F}^{n}_0[O_p]=\\
    %&=\frac{1}{Z_0(E)}\int {\rm d}y\,\frac{e^{yE}}{2\pi i}\,y^{-(N+n)}\int{\rm d}\tilde{\mathbf{q}}{\rm d}\tilde{\mathbf{p}}\, e^{-H_0(\tilde{\mathbf{q}},\tilde{\mathbf{p}})}\tilde{p}_j^{2n}=\\
    %&= \frac{E^n}{T^{n}} \frac{\Gamma(N)}{\Gamma(N+n)}\langle p_j^{2n}\rangle_{c}\\
    &=\frac{N^n\Gamma(N)}{\Gamma(N+n)}\langle p_j^{2n}\rangle_{c}\,;
\end{align}
and 

\begin{align}
    \langle q_j^{2n}\rangle %&=\frac{1}{Z_0(E)}\int {\rm d}y\,\frac{e^{yE}}{2\pi i}\int{\rm d}\mathbf{q}{\rm d}\mathbf{p}\, e^{-yH_0(\mathbf{q},\mathbf{p})}\,q_j^{2n}=\\
    &=\mathcal{F}^{n}_0[O_q]=\\
    %&=\frac{1}{Z_0(E)}\int {\rm d}y\,\frac{e^{yE}}{2\pi i}\,y^{-(N+n)}\int{\rm d}\tilde{\mathbf{q}}{\rm d}\tilde{\mathbf{p}}\, e^{-H_0(\tilde{\mathbf{q}},\tilde{\mathbf{p}})}\tilde{q}_j^{2n}=\\
    %&= \frac{E^n}{T^{n}} \frac{\Gamma(N)}{\Gamma(N+n)}\langle q_j^{2n}\rangle_{c}\\
    &=\frac{N^n \Gamma(N)}{\Gamma(N+n)}\langle q_j^{2n}\rangle_{c}\,.
\end{align}
\end{comment}
From the above formulas, it is immediate to verify that 
\begin{align}
    &n=1\implies\begin{cases}
            &\langle q_j^{2}\rangle= \langle q_j^{2}\rangle_{c}\\ &\langle p_j^{2}\rangle=\langle p_j^{2}\rangle_{c}
        \end{cases}\\
        &n=2\implies\begin{cases}
            &\langle q_j^{4}\rangle= \frac{N}{N+1}\langle q_j^{4}\rangle_{c}\\ &\langle p_j^{4}\rangle=\frac{N}{N+1}\langle p_j^{4}\rangle_{c}
        \end{cases}
\end{align}

\subsubsection{First-order correction}
Let us now focus on the nonlinear case. Assuming that the (adimensional) parameter $\beta$ is small, the partition function $Z(E)$ can be computed perturbatively in powers of $\beta$. 
More precisely, the partition function $Z(E)$ takes the form 
\begin{subequations}
    \begin{align}
    Z(E)&=\int {\rm d}y\,\frac{e^{yE}}{2\pi i}\int{\rm d}\mathbf{q}{\rm d}\mathbf{p}\, e^{-yH(\mathbf{q},\mathbf{p})}=\\
    &=\int {\rm d}y\,\frac{e^{yE}}{2\pi i}\int{\rm d}\mathbf{q}{\rm d}\mathbf{p}\, e^{-yH_0(\mathbf{q},\mathbf{p})-y\beta H_4(\mathbf{q})}=\\
    &=Z_0(E)\left[1-\frac{\beta}{Z_0(E)}\int {\rm d}y\,\frac{y\, e^{yE}}{2\pi i}  \int{\rm d}\mathbf{q}{\rm d}\mathbf{p}\, e^{-yH_0(\mathbf{q},\mathbf{p})} H_4(\mathbf{q})\right]+O(\beta^2)=\\
    %&=Z_0(E)\left[1-\frac{\beta}{Z_0(E)}\int {\rm d}y\,\frac{ e^{yE}}{2\pi i} y^{-(N+1)}  \int{\rm d}\tilde{\mathbf{q}}{\rm d}\tilde{\mathbf{p}}\, e^{-H_0(\tilde{\mathbf{q}},\tilde{\mathbf{p}})}H_4(\tilde{\mathbf{q}})\right]+O(\beta^2)=\\
    %&=Z_0(E)\left(1-\beta \frac{E}{T^2}\frac{\Gamma(N)}{\Gamma(N+1)}\langle  H_4(\mathbf{q})\rangle_{c}\right)+O(\beta^2)\\
    %&=Z_0(E)\left(1-\beta \frac{\langle  H_4(\mathbf{q})\rangle_{c}}{T}\right)+O(\beta^2)\,.
    &=Z_0(E)\left(1-\beta \mathcal{F}_1^{2}[H_4]\right)+O(\beta^2)%\\
    %&=Z_0(E)\left(1-\beta \frac{N}{E}\frac{N \Gamma(N)}{\Gamma(N+1)}\langle  H_4(\mathbf{q})\rangle_{c}\right)+O(\beta^2)\\
    %&=Z_0(E)\left(1-\beta \frac{N}{E}\langle  H_4(\mathbf{q})\rangle_{c}\right)+O(\beta^2)\,
\end{align}
\end{subequations}
At leading order in $\beta$, the inverse of $Z(E)$ reads
\begin{equation}
    %\frac{1}{Z(E)}=\frac{1}{Z_0(E)}\left(1+\beta \frac{\langle  H_4(\mathbf{q})\rangle_{c}}{T}\right)+O(\beta^2)\,.
    %\frac{1}{Z(E)}=\frac{1}{Z_0(E)}\left(1+\frac{\beta N}{E}\langle  H_4(\mathbf{q})\rangle_{c}\right)+O(\beta^2)\,.
    \frac{1}{Z(E)}=\frac{1}{Z_0(E)}\left(1+\beta\mathcal{F}_1^{2}[H_4]\right)+O(\beta^2)\,.
\end{equation}
%At this point, by introducing $O_{pH}\equiv p_j^{2n}H_4(\mathbf{q})$ and $O_{qH}\equiv q_j^{2n}H_4(\mathbf{q})$, the microcanonical averages of $\langle p_j^{2n}\rangle$ and $\langle q_j^{2n}\rangle$ can be written at order $\beta$ as
At this point, %by introducing $\hat{O}_{H}\equiv p_j^{2n}H_4(\mathbf{q})$ and $O_{qH}\equiv q_j^{2n}H_4(\mathbf{q})$, 
the microcanonical average $\langle O_{2n}(\mathbf{q},\mathbf{p})\rangle$ can be computed as
%averages of $\langle p_j^{2n}\rangle$ and $\langle q_j^{2n}\rangle$ can be written at order $\beta$ as
\begin{subequations}
\begin{align}
    &\langle O_{2n}(\mathbf{q},\mathbf{p})\rangle =\frac{1}{Z(E)}\int {\rm d}y\,\frac{e^{yE}}{2\pi i}\int{\rm d}\mathbf{q}{\rm d}\mathbf{p}\, e^{-yH(\mathbf{q},\mathbf{p})}\,O_{2n}(\mathbf{q},\mathbf{p})=\\
    &=\frac{\left(1+\beta\mathcal{F}_1^{2}[H_4]\right)}{Z_0(E)}\int {\rm d}y\,\frac{e^{yE}}{2\pi i}\int{\rm d}\mathbf{q}{\rm d}\mathbf{p}\, e^{-yH_0(\mathbf{q},\mathbf{p})}\,O_{2n}(\mathbf{q},\mathbf{p})(1-\beta y  H_4(\mathbf{q}))+O(\beta^2)=\\
    &=\left(1+\beta\mathcal{F}_1^{2}[H_4]\right)\left(\mathcal{F}_0^n[O]-\beta \mathcal{F}_1^{n+2}[OH_4]\right)+O(\beta^2)=\\
    &=\mathcal{F}_0^n[O]+\beta\left(\mathcal{F}_1^{2}[H_4]\mathcal{F}_0^n[O]- \mathcal{F}_1^{n+2}[OH_4]\right)+O(\beta^2)%=\\
    %&=\frac{N^n\Gamma(N)}{\Gamma(N+n)}\langle O_{2n}(\mathbf{q},\mathbf{p})\rangle_{c}+\beta\left(\frac{N}{E}\frac{N^{n+1}\Gamma(N)}{\Gamma(N+n+1)}\langle p_j^{2n} H_4(\mathbf{q})\rangle_{c}\right)+O(\beta^2)\\
    %&=\frac{N^n\Gamma(N)}{\Gamma(N+n)}\langle p_j^{2n}\rangle_{c}+\beta\frac{N}{E}\frac{N^{n}\Gamma(N)}{\Gamma(N+n)}\langle p_j^{2n}\rangle_{c}\langle  H_4(\mathbf{q})\rangle_{c}\left(1-\frac{N}{N+n}\right)+O(\beta^2)\\
    %&=\frac{N^n\Gamma(N)}{\Gamma(N+n)}\langle p_j^{2n}\rangle_{c}+\beta\frac{N}{E}\frac{N^{n}\Gamma(N)}{\Gamma(N+n)}\frac{n}{N+n}\langle p_j^{2n}\rangle_{c}\langle  H_4(\mathbf{q})\rangle_{c}+O(\beta^2)\,;
\end{align}
\end{subequations}
By taking into account the expression of $\mathcal{F}_m^n[\cdot]$, $\langle O_{2n}(\mathbf{q},\mathbf{p})\rangle$ can be rewritten as
\begin{subequations}
   \begin{align}
    \langle O_{2n}\rangle &=\mathcal{F}_0^n[O]+\beta\left(\mathcal{F}_1^{2}[H_4]\mathcal{F}_0^n[O]- \mathcal{F}_1^{n+2}[OH_4]\right)+O(\beta^2)=\\
    &=\frac{N^n\Gamma(N)}{\Gamma(N+n)}\langle O_{2n}\rangle_{c}+\beta\left(\frac{N^{n} \Gamma(N)}{\Gamma(N+n)}\frac{N}{E}\langle  H_4\rangle_{c}\langle O_{2n}\rangle_{c}-\frac{N}{E}\frac{N^{n+1}\Gamma(N)}{\Gamma(N+n+1)}\langle O_{2n} H_4\rangle_{c}\right)+O(\beta^2)\\
    &=\frac{N^n\Gamma(N)}{\Gamma(N+n)}\langle O_{2n}\rangle_{c}+\beta\frac{N}{E}\frac{N^{n}\Gamma(N)}{\Gamma(N+n+1)}\left[(N+n)\langle  H_4\rangle_{c}\langle O_{2n}\rangle_{c}-N\langle O_{2n} H_4\rangle_{c}\right]+O(\beta^2)\\
    &=\frac{N^n\Gamma(N)}{\Gamma(N+n)}\langle O_{2n}\rangle_{c}+\beta\frac{N}{E}\frac{N^{n}\Gamma(N)}{\Gamma(N+n+1)}\left[N\left(\langle  H_4\rangle_{c}\langle O_{2n}\rangle_{c}-\langle O_{2n} H_4\rangle_{c}\right)+n\langle  H_4\rangle_{c}\langle O_{2n}\rangle_{c}\right]+O(\beta^2)\,.
    %&=\frac{N^n\Gamma(N)}{\Gamma(N+n)}\langle p_j^{2n}\rangle_{c}+\beta\frac{N}{E}\frac{N^{n}\Gamma(N)}{\Gamma(N+n)}\frac{n}{N+n}\langle p_j^{2n}\rangle_{c}\langle  H_4(\mathbf{q})\rangle_{c}+O(\beta^2)\,;
\end{align} 
\end{subequations}

To compute $\langle E_j^{os}\rangle=\frac{1}{2}\left(\langle p_j^{2}\rangle+k_R\langle q_j^{2}\rangle\right)$ we need to estimate the above expression for the observables $p_j^{2}$ and $q_j^{2}$ ($n=1$). By nothing that $\langle  H_4(\mathbf{q}) f_{2n}(\mathbf{p})\rangle_{c}=\langle  H_4(\mathbf{q})\rangle_{c}\langle f_{2n}(\mathbf{p})\rangle_{c}$, one obtains the following expressions
\begin{align}
    &\langle p_j^{2}\rangle=\langle p_j^{2}\rangle_{c}+\frac{\beta N}{E}\frac{\langle p_j^{2}\rangle_{c}\langle  H_4(\mathbf{q})\rangle_{c}}{N+1}+O(\beta^2)\label{eq:p2-correction}\\
    &\langle q_j^{2}\rangle=\langle q_j^{2}\rangle_{c}+\frac{\beta N}{E(N+1)}\left[N\left(\langle q_j^{2}\rangle_{c}\langle  H_4(\mathbf{q})\rangle_{c}-\langle q_j^{2}  H_4(\mathbf{q})\rangle_{c}\right)+\langle q_j^{2}\rangle_{c}\langle  H_4(\mathbf{q})\rangle_{c}\right]+O(\beta^2)\label{eq:q2-correction}
\end{align}
which combined together lead to
\begin{subequations}
\begin{align}
    &\langle E_j^{os}\rangle=\frac{1}{2}\left(\langle p_j^{2}\rangle+k_R\langle q_j^{2}\rangle\right)=\\
    %&=\frac{1}{2}\left(\langle p_j^{2}\rangle_{c}+k_R\langle q_j^{2}\rangle_{c}\right)-\frac{\beta N}{2E}\left[k_R\left(\langle q_j^{2}  H_4(\mathbf{q})\rangle_{c}-\langle q_j^{2}\rangle_{c}\langle  H_4(\mathbf{q})\rangle_{c}\right)-\frac{\langle p_j^{2}\rangle_{c}\langle  H_4(\mathbf{q})\rangle_{c}+k_R\langle q_j^{2}  H_4(\mathbf{q})\rangle_{c}}{N+1}\right]+O(\beta^2)\\
    %&=\frac{E}{2 N}\left(1+k_R \mathsf{M}^{-1}_{jj}\right)-\frac{\beta}{2}\left[k_R\frac{N\langle q_j^{2}  H_4(\mathbf{q})\rangle_{c}}{E}-k_R \mathsf{M}^{-1}_{jj}\langle  H_4(\mathbf{q})\rangle_{c}-\frac{\langle  H_4(\mathbf{q})\rangle_{c}}{N+1}\left(1+\frac{k_R N}{E}\frac{\langle q_j^{2}  H_4(\mathbf{q})\rangle_{c}}{\langle  H_4(\mathbf{q})\rangle_{c}}\right)\right]+O(\beta^2)\\
    &=\frac{1}{2}\left(\langle p_j^{2}\rangle_{c}+k_R\langle q_j^{2}\rangle_{c}\right)+\frac{\beta N}{2(N+1)E}\left[k_RN\left(\langle q_j^{2}\rangle_{c}\langle  H_4(\mathbf{q})\rangle_{c}-\langle q_j^{2}  H_4(\mathbf{q})\rangle_{c}\right)+\frac{E}{N}\left(\langle p_j^{2}\rangle_{c}+k_R\langle q_j^{2}\rangle_{c}\right)\langle  H_4(\mathbf{q})\rangle_{c}\right]+O(\beta^2)\\
    &=\frac{E}{2 N}\left(1+k_R \mathsf{M}^{-1}_{jj}\right)+\frac{\beta N}{2(N+1)E}\left[k_RN\left(\langle q_j^{2}\rangle_{c}\langle  H_4(\mathbf{q})\rangle_{c}-\langle q_j^{2}  H_4(\mathbf{q})\rangle_{c}\right)+\frac{E}{N}(1+k_R \mathsf{M}^{-1}_{jj})\langle  H_4(\mathbf{q})\rangle_{c}\right]+O(\beta^2)\\%\,.
    &=\frac{E\left(1+k_R \mathsf{M}^{-1}_{jj}\right)}{2 N}+\frac{\beta N \left[k_RN\left(\langle q_j^{2}\rangle_{c}\langle  H_4(\mathbf{q})\rangle_{c}-\langle q_j^{2}  H_4(\mathbf{q})\rangle_{c}\right)+\frac{E}{N}(1+k_R \mathsf{M}^{-1}_{jj})\langle  H_4(\mathbf{q})\rangle_{c}\right]}{2(N+1)E}+O(\beta^2)\,.
    %&=\frac{E}{2 N}\left(1+k_R \mathsf{M}^{-1}_{jj}\right)+\frac{\beta}{2(N+1)}\left[\frac{k_RN^2}{E}\left(\langle q_j^{2}\rangle_{c}\langle  H_4(\mathbf{q})\rangle_{c}-\langle q_j^{2}  H_4(\mathbf{q})\rangle_{c}\right)+(1+k_R \mathsf{M}^{-1}_{jj})\langle  H_4(\mathbf{q})\rangle_{c}\right]+O(\beta^2)\,.
\end{align}    
\end{subequations}

\subsubsection{Canonical expectation values}
In this section we provide explicit expressions for the canonical expectation values appearing in Eq.~\eqref{eq:eos_beta}. 
The expectation values  $\langle O(\mathbf{q},\mathbf{p})\rangle_{c}$ over the canonical ensemble at temperature $T=\frac{E}{N}$ are defined as
\begin{equation}
    \langle O(\mathbf{q},\mathbf{p})\rangle_{c}=\frac{\int{\rm d}\mathbf{q}{\rm d}\mathbf{p}\, e^{-H_0(\mathbf{q},\mathbf{p})/T} O(\mathbf{q},\mathbf{p})}{\int{\rm d}\mathbf{q}{\rm d}\mathbf{p}\, e^{-H_0(\mathbf{q},\mathbf{p})/T}}\,.
\end{equation}
%$\rho_\text{can}(T)= \frac{e^{- H_{0}(\mathbf{q},\mathbf{p})/T}}{Z}$
Since $H_0(\mathbf{q},\mathbf{p})$ is separable in $\mathbf{q}$ and $\mathbf{p}$, for every observable $O(\mathbf{q},\mathbf{p})=O_q(\mathbf{q})O_p(\mathbf{p})$ one has
\begin{equation}
    \langle O(\mathbf{q},\mathbf{p})\rangle_{c}=\langle O_q(\mathbf{q})\rangle_{c}\langle O_p(\mathbf{p}) \rangle_{c}\,.
\end{equation}
Therefore, the only terms that need to be computed are $\langle  H_4(\mathbf{q})\rangle_{c}$ and $\langle q_j^{2}  H_4(\mathbf{q})\rangle_{c}$, both involving only polynomials in $\mathbf{q}$. By noting that the canonical density $\rho_\text{can}(T)$ is Gaussian %and that both $\langle  H_4(\mathbf{q})\rangle_{c}$ and $\langle q_j^{2}  H_4(\mathbf{q})\rangle_{c}$ only involve polynomials in $\mathbf{q}$, 
the expectation values can be computed using the Wick formula
\begin{align}
\langle q_{j_1} \cdots q_{j_n} \rangle_{c}
=
\begin{cases}
0, & n  \text{ odd;}\\
\displaystyle
\sum_{\mathcal{P}}
\;\prod_{(i,l)\in \mathcal{P}}
\langle q_{i} q_{l} \rangle_{c}, & n \text{ even;}
\end{cases}
\end{align}
where $\mathcal{P}$ represents the space of all possible partitions in pairs and the expectation value $\langle q_{i} q_{l} \rangle_{c}$ reads 
\begin{equation}
    \langle q_{i} q_{l} \rangle_{c}=T \mathsf{M}^{-1}_{il}=\frac{E}{N}\mathsf{M}^{-1}_{il}\,.
\end{equation}
Let us also note that 
\begin{equation}
    \langle q_j^{2}  H_4(\mathbf{q})\rangle_{c}=\langle q_j^{2} \rangle_{c}\langle H_4(\mathbf{q})\rangle_{c}+\left(\langle q_j^{2}  H_4(\mathbf{q})\rangle_{c}-\langle q_j^{2} \rangle_{c}\langle H_4(\mathbf{q})\rangle_{c}\right)
\end{equation}
and, moreover, that the only  partitions $\mathcal{P'}$ contributing to the term in the bracket are those such that the $q_j$ terms are not coupled with terms coming from $H_4$ (the ``connected'' contributions).
%that do not contain the pair $(j,j)$ contribute to the term in the bracket. 
By considering the explicit expression for $H_4(\mathbf{q})$
\begin{equation}
    H_4(\mathbf{q})=\frac{1}{4}\sum_{n=0}^N\cbr{q_n-q_{n+1}}^4=\frac{1}{4}\sum_{n=0}^N\cbr{q_n^4+q_{n+1}^4-4q_n^3q_{n+1}-4q_nq_{n+1}^3+6q_n^2q_{n+1}^2}\,,
\end{equation}  
%and by taking into account the expression for $\langle q_{i} q_{l} \rangle_{c}$, i.e.  

one finally obtains
\begin{align*}
    \langle H_4(\mathbf{q})\rangle_{c}&=\frac{E^2}{4 N^2}\sum_{n=0}^N\left[3\left(\mathsf{M}^{-1}_{nn}\right)^2+3\left(\mathsf{M}^{-1}_{n+1,n+1}\right)^2-4\left(3\mathsf{M}^{-1}_{n+1,n+1}\mathsf{M}^{-1}_{n+1,n}+3\mathsf{M}^{-1}_{n,n}\mathsf{M}^{-1}_{n,n+1}\right)\right]+\nonumber\\
    &+\frac{E^2}{4 N^2}\sum_{n=0}^N\left[6\left(\mathsf{M}^{-1}_{n+1,n+1}\mathsf{M}^{-1}_{n,n}+2\left(\mathsf{M}^{-1}_{n,n+1}\right)^2\right)\right]=\\
    &=\frac{3 E^2}{4 N^2}\sum_{n=0}^N \left[ \left(\mathsf{M}^{-1}_{nn}\right)^2 +\left(\mathsf{M}^{-1}_{n+1,n+1}\right)^2-4\mathsf{M}^{-1}_{n,n+1}\left(\mathsf{M}^{-1}_{n+1,n+1}+\mathsf{M}^{-1}_{nn}-\mathsf{M}^{-1}_{n,n+1}\right)+2\mathsf{M}^{-1}_{n+1,n+1}\mathsf{M}^{-1}_{nn}\right]=\\
    &=\frac{3 E^2}{4 N^2}\sum_{n=0}^N \left[\mathsf{M}^{-1}_{n+1,n+1}+\mathsf{M}^{-1}_{nn}-2\mathsf{M}^{-1}_{n,n+1}\right]^2\,,
\end{align*}
and
\begin{align*}
    \langle q_j^{2}  H_4(\mathbf{q})\rangle_{c}-\langle q_j^{2} \rangle_{c}\langle H_4(\mathbf{q})\rangle_{c}&=\frac{E^3}{4 N^3}\sum_{n=0}^N 12\left[\mathsf{M}^{-1}_{n+1,n+1}\left(\mathsf{M}^{-1}_{j,n+1}\right)^2+\mathsf{M}^{-1}_{nn}\left(\mathsf{M}^{-1}_{jn}\right)^2\right]+\nonumber\\
    &+\frac{E^3}{4 N^3}\sum_{n=0}^N 12\left[\mathsf{M}^{-1}_{nn}\left(\mathsf{M}^{-1}_{j,n+1}\right)^2+\mathsf{M}^{-1}_{n+1,n+1}\left(\mathsf{M}^{-1}_{jn}\right)^2+2\mathsf{M}^{-1}_{n,n+1}\mathsf{M}^{-1}_{j,n+1}\mathsf{M}^{-1}_{jn}\right]-\nonumber\\
    &-\frac{E^3}{4 N^3}\sum_{n=0}^N 12\left[\mathsf{M}^{-1}_{j,n+1}\mathsf{M}^{-1}_{jn}\left(\mathsf{M}^{-1}_{n+1,n+1}+\mathsf{M}^{-1}_{nn}\right)+2\mathsf{M}^{-1}_{n,n+1}\left(\left(\mathsf{M}^{-1}_{j,n+1}\right)^2+\left(\mathsf{M}^{-1}_{jn}\right)^2\right)\right]=\\
    &=\frac{3 E^3}{ N^3}\sum_{n=0}^N\left[\left(\mathsf{M}^{-1}_{n+1,n+1}+\mathsf{M}^{-1}_{nn}-2\mathsf{M}^{-1}_{n,n+1}\right)\left(\left(\mathsf{M}^{-1}_{j,n+1}\right)^2+\left(\mathsf{M}^{-1}_{jn}\right)^2-\mathsf{M}^{-1}_{j,n+1}\mathsf{M}^{-1}_{jn}\right)\right]\,.
\end{align*}

\bibliographystyle{apsrev4-1}
\bibliography{biblio}

\end{document}